\documentclass[conference]{IEEEtran}
\IEEEoverridecommandlockouts
\usepackage{cite}
\usepackage{amsmath,amssymb,amsfonts}
\usepackage{algorithmic}
\usepackage{graphicx}
\usepackage{textcomp}
\usepackage{xcolor}

\usepackage{booktabs}
\usepackage{tabularx}
\usepackage{array}
\usepackage{makecell}
\usepackage{subcaption}
\usepackage{adjustbox}
\usepackage{multirow}
\usepackage{array}
\usepackage{float}
\usepackage[hidelinks]{hyperref}

\def\BibTeX{{\rm B\kern-.05em{\sc i\kern-.025em b}\kern-.08em
    T\kern-.1667em\lower.7ex\hbox{E}\kern-.125emX}}
\begin{document}

\title{MAGE: Meta-path Adaptive Graph Evidence for Transferable Multimodal Graph Learning

}

\author{
\IEEEauthorblockN{
Haiyang Jiang\textsuperscript{1},
Tong Chen\textsuperscript{1},
Xinyi Gao\textsuperscript{1},
Hao Wang\textsuperscript{2},
Hongzhi Yin\textsuperscript{1}
}
\IEEEauthorblockA{
\textsuperscript{1}\textit{The University of Queensland, Brisbane, Australia},
\textsuperscript{2}\textit{Alibaba Group, Beijing, China}\\
\{\href{mailto:haiyang.jiang@uq.edu.au}{haiyang.jiang},
\href{mailto:tong.chen@uq.edu.au}{tong.chen},
\href{mailto:h.yin1@uq.edu.au}{h.yin1}\}@uq.edu.au\\
\href{mailto:xinyi_gao@outlook.com}{xinyi\_gao@outlook.com},
\href{mailto:cashenry@126.com}{cashenry@126.com}
}
}




\maketitle

\begin{abstract}
Multimodal graphs integrate textual and visual attributes with relational structures, providing a rich representation of complex real-world data.
To enable the broad applicability and reuse of graph models, it is crucial for models trained on source graphs to generalize to unseen graph domains.
However, this is challenging because cross-modal interactions among neighbors and the neighboring information needed for prediction can vary across target nodes and graph domains.
This gives rise to two key challenges: \emph{adaptive cross-modal neighborhood fusion}, which captures and integrates cross-modal interactions among neighbors to accommodate variations across nodes and domains; and \emph{target-specific evidence selection}, which identifies the neighboring textual and visual information needed for each target node to support LLM-based prediction.
In light of these challenges, we propose MAGE (\underline{M}eta-path \underline{A}daptive \underline{G}raph \underline{E}vidence Prompting), a framework for transferable multimodal graph learning.
MAGE represents textual and visual node views separately in a heterogeneous graph and introduces two complementary components.
Meta-path Adaptive Multimodal Propagation (MAMP) adaptively aggregates neighborhood information along heterogeneous meta-paths and compresses it into graph-aware soft tokens.
Evidence-Guided Instruction Tuning (EGIT) selects relevant neighboring evidence for each query and incorporates its graph-aware representations into instructions for a frozen multimodal LLM.
Together, these components support transfer from multiple source graphs to unseen target graphs without target-domain parameter updates.
Experiments with three source graphs and four unseen target graphs demonstrate strong cross-domain transfer performance on node classification and link prediction, without requiring parameter updates on the target domains.
\end{abstract}


\section{Introduction}
\label{sec:introduction}
Web applications commonly integrate textual, visual, and relational information, as seen in social media posts linked by user interactions~\cite{chen2025birds,cheng2024retrieval,shu2020fakenewsnet,qi2012community} and e-commerce products connected through co-purchases or co-interactions~\cite{deng2023construction,he2024firzen,mcauley2015image,zhang2021mining}.
These data can be naturally represented as multimodal graphs, where nodes contain textual and visual attributes and edges capture their underlying relationships~\cite{he2025unigraph2,fan2026mlaga}.
Understanding multimodal graphs requires not only interpreting the textual and visual content of individual nodes but also incorporating neighborhood information through graph topology~\cite{zhou2025multimodal,zhang2025multiple}. 
However, existing methods aggregate textual and visual information separately within each modality and fuse them only afterward, failing to fully capture cross-modal interactions during neighborhood aggregation~\cite{li2026modalities,zhang2025multiple,peng2025modalities}. Consequently, their generalization to unseen graph domains remains limited.
For example, the text of a social media post may express the author's opinions or intentions, while its accompanying image provides visual context that is not explicitly conveyed in the text.
Consequently, a target node may benefit more from the textual information of one neighbor and the visual information of another, making effective multimodal neighborhood aggregation challenging.
Nevertheless, the relative contributions of neighboring modalities can vary across domains. Methods that learn simplistic modality fusion mechanisms may therefore struggle to transfer knowledge to unseen domains~\cite{li2026modalities,zhang2025multiple,peng2025modalities,li2026choe,zhu2021transfer,yarici2026mer,planamente2024relative,fan2024cross}. This motivates adaptive multimodal neighborhood aggregation for cross-domain transfer.

\begin{figure}[t]
    \centering
    \includegraphics[width=\columnwidth]
    {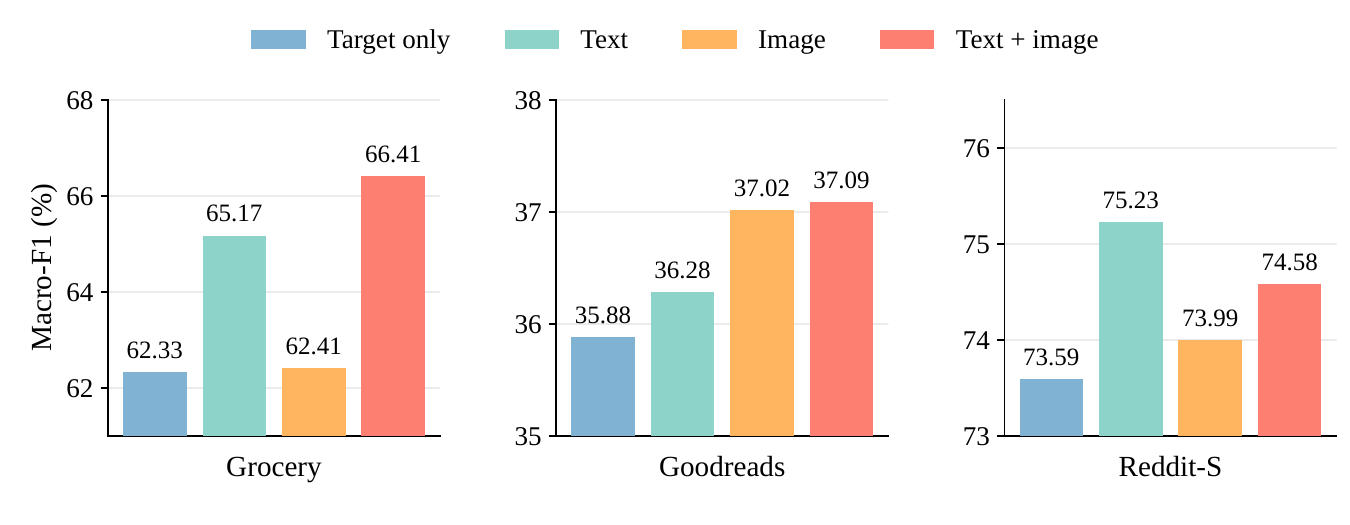}
    \caption{Neighbor-modality utility on three datasets.
    A frozen Qwen2.5-VL-7B-Instruct model is evaluated on
    500 fixed test nodes per dataset with up to three
    training-set neighbors, without neighbor labels.
    Target only uses no neighbors; Text, Image, and
    Text + image specify the available neighbor modalities.
    Target text and image are retained in all conditions.
    Bars show Macro-F1 point estimates.}
    \label{fig:neighbor-modality-motivation}
\end{figure}

Although recent multimodal large language models (LLMs) have demonstrated remarkable capabilities in visual understanding and multimodal reasoning~\cite{liu2023visual,li2023blip,dai2023instructblip,alayrac2022flamingo}, semantic alignment between text and images~\cite{radford2021learning,li2023blip,zhai2023sigmoid} alone does not determine how information should be exchanged across graph nodes. 
In particular, neighboring text and images can provide complementary contextual cues,
yet their usefulness for target-node prediction varies.
To investigate this issue, we conduct a controlled node classification study using a frozen multimodal LLM. For each test node, we select a fixed set of neighbors and evaluate three input settings: (1) neighbor text only, (2) neighbor images only, and (3) both neighbor text and images, to examine how different neighboring modalities affect prediction performance.
As shown in Figure~\ref{fig:neighbor-modality-motivation}, the results highlight two key findings.
First, combining neighboring text and images does not always yield the best performance: text-only neighbors achieve the best results on Reddit-S.
Second, the contribution of each neighboring modality varies across datasets.
Text-only neighbors outperform image-only neighbors on Grocery (65.17\% versus 62.41\%), whereas image-only neighbors outperform text-only neighbors on Goodreads (37.02\% versus 36.28\%).
These observations suggest that the relative utility of neighboring modalities can vary across datasets, and simply providing both modalities does not necessarily yield the best prediction.

This variability complicates cross-domain transfer, as neighborhood aggregation strategies learned from source graphs may not generalize to unseen domains with different textual, visual, and structural characteristics. Beyond neighborhood aggregation, LLM-based prediction also requires careful evidence selection: different target nodes may require different neighboring information, while including all available text and images can introduce irrelevant or redundant information that obscures useful contextual cues~\cite{li2024similarity,he2024g,mortaheb2025re}. Moreover, evidence selection strategies learned from source graphs may fail to identify the information needed for prediction in unseen domains.
These considerations give rise to two key challenges. First, \emph{adaptive cross-modal neighborhood fusion} requires capturing cross-modal interactions among neighbors while adapting their contributions across nodes and domains. Second, \emph{target-specific evidence selection} requires identifying informative neighboring text and images to construct target-specific LLM prompts that support cross-domain generalization.

To this end, we propose \textbf{MAGE} (\textbf{M}eta-path \textbf{A}daptive \textbf{G}raph \textbf{E}vidence Prompting), an adaptive evidence prompting framework for transferable multimodal graph learning. MAGE first represents the textual and visual modalities of each original node as two distinct modality-specific nodes. It then expands the original edges into intra-modal and cross-modal connections, producing a heterogeneous graph that enables interactions between different modalities across neighboring nodes.
Building on this graph, MAGE comprises two components. 
To address the first challenge, \textbf{Meta-path Adaptive Multimodal Propagation (MAMP)} integrates multimodal neighborhood information through heterogeneous meta-path message passing to obtain path-specific representations. A Meta-path Semantic Prompt Router (MSPR) adaptively fuses these representations for each node and modality, enabling flexible aggregation of different neighboring modalities across domains. Finally, modality-specific Q-Formers compress the contextualized token sequences into compact graph-aware soft tokens for efficient LLM reasoning.
For the second challenge, \textbf{Evidence-Guided Instruction Tuning (EGIT)} selects relevant neighboring evidence for each target query and incorporates its representations into task instructions. Each evidence item carries graph-aware representations that adaptively integrate textual and visual information from its constituent nodes' neighborhoods, providing additional contextual cues beyond the target's representation. This connects adaptive heterogeneous propagation with evidence-based prediction, enabling the transfer of knowledge from multiple source graphs to unseen graph domains.

Our main contributions are summarized as follows:
\begin{itemize}
    \item 
    We identify the varying contributions of neighboring modalities as an important challenge for cross-domain multimodal knowledge transfer.
    Through a controlled study, we show that the relative utility of neighboring text and images varies across datasets, motivating explicit modeling of their contributions beyond node-level modality alignment.

    \item We propose MAGE, whose MAMP component explicitly models cross-node and cross-modal dependencies through a modality-typed heterogeneous graph and path-specific propagation. Its Meta-path Semantic Prompt Router combines node- and neighborhood-conditioned routing with path-specific semantic prompt matching to adaptively fuse propagation patterns.

    \item We introduce EGIT to select relevant neighboring evidence and incorporate its graph-aware representations into task instructions. These representations encode adaptively aggregated textual and visual information from the evidence nodes' own neighborhoods, providing supplementary context beyond the target representation for prediction on unseen graphs.

    \item We train MAGE on three source datasets and evaluate its
cross-domain transfer performance on four unseen target datasets
for both node classification and link prediction.
The results demonstrate strong transfer performance and
improvements over competitive baselines across both tasks.
\end{itemize}

\section{Related Work}
\label{sec:related_work}

\subsection{Large Language Models for Graphs}
\label{sec:llms_for_graphs}
Recent studies have explored integrating graph structural information with large language models (LLMs) for graph understanding and prediction~\cite{wang2024llms,qiao2025login,wang2024instructgraph,he2024harnessing}.
GraphPrompter~\cite{liu2024can} encodes local subgraphs with a GNN and projects their representations into soft prompts for a frozen LLM.
GraphGPT~\cite{tang2024graphgpt} combines text--graph grounding with self-supervised and task-specific instruction tuning to incorporate structural knowledge into LLMs.
LLaGA~\cite{chen2024llaga} organizes graph neighborhoods into structure-aware sequences using Neighborhood Detail and Hop-Field Overview templates,
and maps these sequences into the LLM token embedding space through a learnable projector.

Beyond graph prompting, several methods investigate transferable graph--language alignment and joint modeling.
GraphTranslator~\cite{zhang2024graphtranslator} bridges pretrained graph models and LLMs through a translation module, supporting both predefined
prediction tasks and open-ended graph understanding.
TEA-GLM~\cite{wang2024llms} aligns GNN representations with LLM token embeddings during pre-training and subsequently learns a linear projector for cross-dataset and cross-task transfer.
GOFA~\cite{kong2025gofa} interleaves GNN layers with a frozen pretrained LLM to combine graph message passing with autoregressive language modeling.
These studies primarily investigate textual attributes and graph structures, whereas our work considers the joint exploitation of textual and visual neighborhood information.

\subsection{Heterogeneous Graph Learning}
Heterogeneous graph learning captures structural and semantic dependencies among different node and relation types~\cite{wang2019heterogeneous,fu2020magnn,hu2020heterogeneous,wang2021self}. PathSim~\cite{sun2011pathsim} uses meta-paths to distinguish relation semantics for similarity search, while ConCH~\cite{li2021leveraging} jointly learns node and meta-path context representations with attention-based fusion for classification. More recent methods improve heterogeneous information aggregation from different perspectives. HGAMLP~\cite{liang2024hgamlp} combines local knowledge extraction, de-redundancy of higher-order meta-path information, and node-adaptive fusion of local and global knowledge. HCAN~\cite{zhang2025effective} incorporates long-range relational context through convolutional subgraph encoding to guide heterogeneous attention. MSGNN~\cite{liu2024msgnn} organizes semantic contexts through schema instances and combines masked modeling with bi-level self-supervision to learn heterogeneous graph representations.
To support transfer across heterogeneous graphs, MUG~\cite{shan2026mug} unifies heterogeneous inputs and employs a shared encoder across meta-path views to capture transferable patterns. MAGE combines adaptive cross-modal neighborhood aggregation with graph-aware evidence to support LLM-based prediction across domains.

\subsection{Multimodal Graph Learning}
\label{sec:multimodal_graph_learning}
Multimodal graph learning integrates textual, visual, and structural information through diverse fusion strategies~\cite{luo2025graph}.
UniGraph2~\cite{he2025unigraph2} averages modality features and applies gated MLP experts followed by graph propagation to learn unified representations.
DiP~\cite{hong2026multimodal} enables cross-modal fusion through proximity-weighted interactions among modality-specific pseudo nodes.
MoMoK~\cite{zhang2025multiple} combines modality-specific experts through relation-guided gating and fuses structural, visual, and textual representations using entity-specific attention.
Beyond graph learning, AMC~\cite{peng2025modalities} adaptively replaces modality tokens and integrates the resulting sequences through self-attention for medical multimodal learning.
MLaGA~\cite{fan2026mlaga} fuses textual and visual tokens through learnable-query cross-attention, supported by structure-aware pre-training.
Mario~\cite{sun2026mario} combines topology-aware attention and cross-modal alignment with adaptive selection of textual, visual, or multimodal instruction views.
Despite these advances in multimodal fusion, explicitly distinguishing and adaptively combining cross-node, cross-modal propagation patterns remains less explored, particularly when the predictive relevance of neighboring modalities varies across graph domains.

Building on these advances, MAGE focuses on explicitly modeling interactions between different neighboring modalities.
It represents the textual and visual views of each node as distinct nodes in a heterogeneous graph.
Meta-path Adaptive Multimodal Propagation adaptively aggregates information along heterogeneous propagation paths and condenses it into graph-aware soft tokens.
Evidence-Guided Instruction Tuning then selects relevant neighboring nodes whose representations capture distinct combinations of textual and visual neighborhood information, providing contextual evidence for prediction under multimodal graph transfer settings.

\section{Problem Formulation}
In this chapter, we first introduce single-domain prediction, then cross-domain prediction. 
A multimodal graph is defined as  
$\mathcal{G} = (\mathcal{V}, \mathcal{E}, \mathcal{I}, \mathcal{T}, \mathcal{Y})$, where $\mathcal{V}$ is the node set, $\mathcal{E}$ is the edge set, $\mathcal{I}$ is the collection of node image features, $\mathcal{T}$ is the set of text features, $\mathcal{Y}$ is the label set. Notably, each node $v_i \in \mathcal{V}$ has an image feature $\mathcal{I}_{v_i} \in \mathcal{I}$ and a text feature $\mathcal{T}_{v_i} \in \mathcal{T}$, and may have a label $\mathcal{Y}_{v_i} \in \mathcal{Y}$. If node $v_i$ is connected to node $v_j$, then $(v_i, v_j) \in \mathcal{E}$. 

\noindent\textbf{Single-domain prediction.}
Given a multimodal graph, the aim is to learn a model $f$ on the training split and make predictions on the test split, fed with both image and text features $( \mathcal{I}, \mathcal{T})$ and structure information $\mathcal{E}$. The prediction task can be node classification: predicting the node label, or edge prediction: estimating whether node $v_i$ and $v_j$ are connected. 

\noindent\textbf{Cross-domain prediction.}
Given source and target multimodal graph collections $\mathcal{D}_s=\{\mathcal{G}_i^s\}_{i=1}^{S}$ and $\mathcal{D}_t=\{\mathcal{G}_j^t\}_{j=1}^{T}$ with
$\mathcal{D}_s\cap\mathcal{D}_t=\emptyset$, we train a shared model $f_{\theta}$ on the labeled source domains by optimizing $\theta^{*}=\arg\min_{\theta}\mathcal{L}_{\mathrm{src}}(f_{\theta})$, where $\theta$ denotes the model parameters and
$\mathcal{L}_{\mathrm{src}}$ is the source-domain training loss. The optimized model then predicts $\widehat{y}_{q}=f_{\theta^{*}}(q;\mathcal{G}_j^t)$
on each unseen target graph $\mathcal{G}_j^t\in\mathcal{D}_t$ without further parameter updates, where $q$ denotes a node for node classification or a node pair for link prediction.

\begin{figure*}[t]
    \centering
    \includegraphics[width=\textwidth]{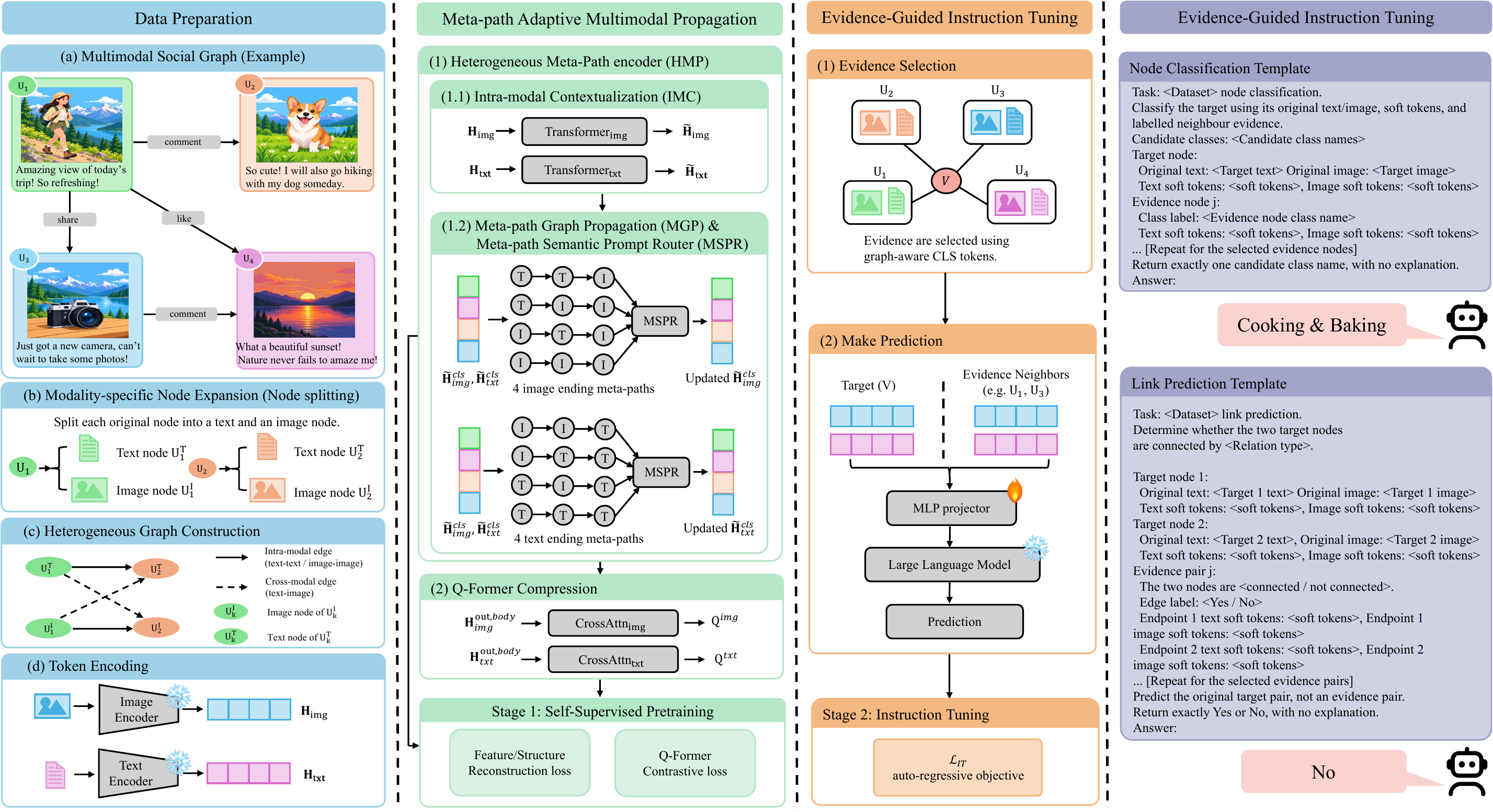}
    \caption{
    Overview of MAGE.
    \textbf{Data preparation:} Each original node is split into text
    and image nodes to construct a modality-typed heterogeneous graph,
    with pretrained encoders providing modality-specific token sequences.
    \textbf{MAMP:} Intra-modal contextualization alternates with
    meta-path propagation and semantic prompt routing. Eight path-specific
    CLS representations are adaptively fused into two modality-specific
    representations and reinserted into the token sequences.
    Q-Formers then compress the contextualized body tokens.
    Stage 1 optimizes these modules using feature reconstruction,
    structure reconstruction, and a contrastive objective that preserves
    modality semantics during compression.
    \textbf{EGIT:} Relevant neighbors are selected as evidence, and
    the target and evidence soft tokens are projected into the LLM
    input space. Stage 2 optimizes only the MLP projectors through
    an autoregressive instruction-tuning loss, while keeping MAMP,
    including its Q-Formers, and the LLM frozen.
    }
    \label{fig:framework}
\end{figure*}

\section{Methodology}
\label{method}
In this section, we introduce the detailed architecture of MAGE, as shown in Fig.~\ref{fig:framework}. 
Firstly, MAGE reorganizes the original graph into a modality-typed heterogeneous graph and encodes each node's text and image into token sequences.
Next, \textbf{Meta-path Adaptive Multimodal Propagation (MAMP)} contextualizes these sequences through alternating intra-modal
attention and adaptive meta-path propagation, followed by Q-Former
compression to obtain compact graph-aware soft tokens.
Then, \textbf{Evidence-Guided Instruction Tuning (EGIT)} selects relevant neighbors as evidence and projects the target
and evidence soft tokens into the LLM input space for prediction.
Training proceeds in two stages: self-supervised pre-training of the
graph representation modules, followed by instruction tuning of the
MLP projectors with the pretrained representation modules and LLM
kept frozen.

\subsection{Data Preparation}
\label{sec:data_preparation}

We prepare the inputs by encoding multimodal node content and
constructing a heterogeneous graph that distinguishes textual and
visual interactions.

\noindent\textbf{Heterogeneous Graph Construction.}
To enable message passing across nodes and modalities, we transform the original multimodal graph into a heterogeneous graph, where modality-specific features are aggregated along meta-paths.
Given the original graph $\mathcal{G}=(\mathcal{V},\mathcal{E})$,
we construct a heterogeneous graph
$\mathcal{G}_{\mathrm{het}}
=(\mathcal{V}_{\mathrm{het}},\mathcal{E}_{\mathrm{het}},
\tau_V,\tau_E)$,
where $\tau_V$ and $\tau_E$ denote the node-type and edge-type
mappings, respectively.

The node-type set is $\mathcal{A}=\{I,T\}$, corresponding to image
and text types.
We represent the image and text modalities of each original node
$v_i$ as two distinct nodes, $v_{I_i}$ and $v_{T_i}$, associated
with $\mathbf{H}_{img,v_i}$ and $\mathbf{H}_{txt,v_i}$, respectively.
Thus, the node set is
$\mathcal{V}_{\mathrm{het}}
=\{v_{M_i}\mid v_i\in\mathcal{V},\,M\in\mathcal{A}\}$,
and the mapping
$\tau_V:\mathcal{V}_{\mathrm{het}}\rightarrow\mathcal{A}$
assigns $\tau_V(v_{M_i})=M$.

For each original edge $(v_i,v_j)\in\mathcal{E}$, we add
$(v_{M_i},v_{M'_j})$ to $\mathcal{E}_{\mathrm{het}}$ for all
$M,M'\in\mathcal{A}$.
For undirected graphs, original edges are considered in both directions.
The edge-type set is
$\mathcal{R}=\{r_{II},r_{IT},r_{TI},r_{TT}\}$,
where $r_{MM'}$ denotes a relation from node type $M$ to node
type $M'$.
The mapping
$\tau_E:\mathcal{E}_{\mathrm{het}}\rightarrow\mathcal{R}$
assigns
$\tau_E((v_{M_i},v_{M'_j}))=r_{MM'}$.
This construction preserves the original neighborhood relationships
while distinguishing intra-modal relations ($r_{II}$ and $r_{TT}$)
from cross-modal relations ($r_{IT}$ and $r_{TI}$).

A two-hop meta-path composes two consecutive relation types
through an intermediate node type.
The resulting graph schema admits eight two-hop meta-paths:
\begin{equation}
\mathcal{P}
=
\left\{
M_1\rightarrow M_2\rightarrow M_3
\mid M_1,M_2,M_3\in\mathcal{A}
\right\}.
\label{eq:meta_path_set}
\end{equation}
Here, the two transitions correspond to relations
$r_{M_1M_2}$ and $r_{M_2M_3}$, respectively.

\noindent\textbf{Token Representations.}
Given the image $\mathcal{I}_{v_i}$ and text $\mathcal{T}_{v_i}$ of
node $v_i$, we use pretrained ViT~\cite{dosovitskiy2020image} and T5~\cite{raffel2020exploring} encoders, denoted by $\phi(\cdot)$
and $g(\cdot)$, to produce modality-specific token representations for each node.
The image sequence
$\mathbf{H}_{img,v_i}=\phi(\mathcal{I}_{v_i})
\in\mathbb{R}^{n_v\times d_i}$
contains a CLS token $\mathbf{h}_{img,v_i}^{\mathrm{CLS}}$
followed by image patch tokens.
For text, we compute the mean of the T5 token embeddings as
$\mathbf{h}_{txt,v_i}^{\mathrm{CLS}}$ and prepend it to the text
token sequence, obtaining
$\mathbf{H}_{txt,v_i}\in\mathbb{R}^{n_t\times d_t}$.
Here, $n_v$ and $n_t$ are the sequence lengths including CLS, and $d_i$ and $d_t$ are the respective feature dimensions.
The first token of each sequence serves as a summary of its modality.

\subsection{Meta-path Adaptive Multimodal Propagation}
\label{HMP}
Given the prepared heterogeneous graph and token sequences, MAMP
adaptively integrates neighborhood information through a heterogeneous
meta-path encoder (HMP) and condenses the resulting sequences using
modality-specific Q-Formers~\cite{li2023blip}.
Then, we design self-supervised objectives to guide the preservation of modality semantics, graph topology, and token information during representation learning.

\subsubsection{Heterogeneous Meta-path Encoder}
As shown in Fig.~\ref{fig:framework}, the HMP encoder repeatedly
applies three modules.
\textit{Intra-modal Contextualization} (IMC) contextualizes each complete modality-specific token sequence.
\textit{Meta-path Graph Propagation} (MGP) then propagates the CLS representations along the eight meta-paths,
and \textit{Meta-path Semantic Prompt Router} (MSPR) fuses the four representations ending in each modality into
a single modality-specific CLS representation.
The two routed CLS representations are reinserted into their respective
token sequences before the next IMC block.
Encoder-block indices are omitted below for clarity.

\noindent\textbf{Intra-modal Contextualization (IMC).}
To capture contextual dependencies within each modality, we apply a
modality-specific Transformer block to the complete token sequence of
node $v_i$:
\begin{equation}
\widetilde{\mathbf{H}}_{m,v_i}
=
\operatorname{Transformer}_{m}
\left(\mathbf{H}_{m,v_i}\right)
\in\mathbb{R}^{n_m\times d},
\label{eq:intra_modal_contextualization}
\end{equation}
where $m\in\{\mathrm{img},\mathrm{txt}\}$ denotes the modality,
$n_m$ is the sequence length including the CLS token
($n_{\mathrm{img}}=n_v$ and $n_{\mathrm{txt}}=n_t$), and $d$ is the
common hidden dimension. Through self-attention, each block updates
the body-token representations and enables the CLS token to aggregate
their information into a contextualized summary of the corresponding
modality.

\noindent\textbf{Meta-path Graph Propagation (MGP).}
We apply graph convolutional networks (GCNs)~\cite{kipf2016semi} along meta-paths to exchange information across nodes and modalities, enriching node representations with multimodal neighborhood context.
We extract the first token of each contextualized sequence as its CLS
representation and stack them across nodes as
$\widetilde{\mathbf{H}}_M^{\mathrm{CLS}}$, $M\in\{T,I\}$.
For each meta-path
$p=(M_1\rightarrow M_2\rightarrow M_3)\in\mathcal{P}$,
we initialize
$\mathbf{H}_p^{(0)}=\widetilde{\mathbf{H}}_{M_1}^{\mathrm{CLS}}$
and perform two-hop propagation:
\begin{equation}
\mathbf{H}_p^{(\ell)}
=
\operatorname{ELU}\Bigl(
\bigl[
\widehat{\mathbf{A}}_{M_\ell\rightarrow M_{\ell+1}}
\mathbf{H}_p^{(\ell-1)}
+
\mathbf{C}\widetilde{\mathbf{H}}_{M_{\ell+1}}^{\mathrm{CLS}}
\bigr]\mathbf{W}_p^{(\ell)}
\Bigr),
\label{eq:meta_path_gnn}
\end{equation}
where $\ell\in\{1,2\}$ and $\widehat{\mathbf{A}}_{M_a\rightarrow M_b}$ is the
degree-normalized adjacency matrix that aggregates messages from
modality $M_a$ to modality $M_b$, excluding self-connections.
The diagonal matrix $\mathbf{C}$ provides degree-normalized
self weights, and $\mathbf{W}_p^{(\ell)}$ is a learnable
transformation specific to each path and hop.
Neighbor messages use representations from the previous hop. The self term uses the same receiving-modality CLS features from the current IMC output at both hops.

Each MGP block produces eight CLS representations per node, with four paths ending in each modality. Let $\mathcal{P}_m$ denote the paths
ending in $m\in\{T,I\}$ and $\mathbf{h}_{v,p}^{m}$ denote the representation
of node $v$ in $\mathbf{H}_p^{(2)}$ for $p\in\mathcal{P}_m$.
These path-specific representations are passed directly to MSPR;
the complete token sequences are not updated until routing is performed.

\begin{table}[t]
\centering
\caption{Shared statistical inputs to the text and image base routers.
Both routers receive the same 5-dimensional node-level descriptor and
6-dimensional local-domain descriptor, but use independent parameters.
Count denotes the number of scalar features.}

\label{tab:router_features}
\small
\setlength{\tabcolsep}{4pt}
\renewcommand{\arraystretch}{1.08}
\begin{tabularx}{\columnwidth}{@{}Xr@{}}
\toprule
\textbf{Feature group} & \textbf{Count} \\
\midrule
\multicolumn{2}{@{}l}{\textit{Node-Level Features (5)}} \\

Node-to-neighborhood-mean cosine similarities:
text--text and image--image
& 2 \\

Text and image neighborhood feature variances
& 2 \\

Log node degree
& 1 \\

\midrule
\multicolumn{2}{@{}l}{\textit{Local-Domain Features (6)}} \\

Mean feature similarities across local edges:
text--text, image--image, and symmetric cross-modal
& 3 \\

Mean center--neighbor cosine similarities:
text--text and image--image
& 2 \\

Local mean log degree
& 1 \\

\bottomrule
\end{tabularx}
\vspace{-0.0cm}
\end{table}

\noindent\textbf{Meta-path Semantic Prompt Router (MSPR).}
\label{sec:mspr}
For each modality-ending branch, MSPR merges the four path-specific representations into one embedding for the corresponding modality node.
MSPR consists of two parts: a base router and a semantic router.
To support cross-domain generalization, the base router uses statistical description features that summarize each node and its local graph context~\cite{kang2026ss,ma2024mixture}.
These summaries reduce its direct dependence on high-dimensional embedding coordinates.
Specifically, we summarize the original text and image CLS representations and local connectivity using a node-level descriptor $\mathbf{r}_v\in\mathbb{R}^{5}$ and a local-domain descriptor $\mathbf{d}_v\in\mathbb{R}^{6}$.


For terminal modality $m\in\{T,I\}$, the base routing score is
\begin{equation}
s_{v,p}^{m,\mathrm{base}}
=
\left[f_{\mathrm{node}}^{m}(\mathbf{r}_v)\right]_p
+
\lambda_{\mathrm{local}}
\left[f_{\mathrm{local}}^{m}(\mathbf{d}_v)\right]_p,
\label{eq:base_router}
\end{equation}
where $f_{\mathrm{node}}^m$ and $f_{\mathrm{local}}^m$ are two-layer MLP,
and $[\cdot]_p$ selects the score for meta-path $p$.
The text and image routers have independent parameters.
Thus, modality specificity arises from the learned routing functions,
rather than from different input descriptors.
The coefficient $\lambda_{\mathrm{local}}$ controls the contribution
of local-domain statistics.



The node-level descriptor characterizes the target node and its relationship with its neighbors.
It includes two cosine similarities between the node's modality representations and the corresponding neighborhood means, text and image neighborhood variances, and log-degree.
These features capture node--neighbor agreement, neighborhood consistency, and connectivity, providing cues for node-specific meta-path weighting.
Both routers receive this same scalar feature.

The local-domain descriptor summarizes feature relationships and connectivity patterns across the local graph context.
It includes feature similarities across local edges, average pairwise similarities between the center and other local nodes, and mean log-degree.
These summaries provide additional routing cues about the surrounding graph environment.
Together, the two descriptors allow the base router to account for both individual node characteristics and local graph patterns when transferring across domains.
All statistics are computed without class labels.
Table~\ref{tab:router_features} lists the shared descriptors.

The base router uses node-level and local-domain statistics to learn routing patterns shared across source domains.
However, these statistical summaries may not fully capture the semantic relevance of each meta-path in an unseen target graph.
We therefore introduce a semantic router to refine the base scores using path-specific representations from the current MGP block.


Each meta-path $p$ is associated with a learnable semantic prompt
$\mathbf{p}_p^m\in\mathbb{R}^{d_p}$.
We first project the path representation $\mathbf{h}_{v,p}^{m}$
into the prompt space using an MLP.
The \emph{semantic score} is then computed as the dot-product
similarity between the projected embedding and its corresponding
prompt vector:
\begin{equation}
s_{v,p}^{m,\mathrm{sem}}
=
\left\langle
\operatorname{GELU}\!\left(
\mathbf{W}_m\mathbf{h}_{v,p}^{m}+\mathbf{b}_m
\right),
\mathbf{p}_p^m
\right\rangle,
\label{eq:prompt_router}
\end{equation}
where $\mathbf{W}_m$ and $\mathbf{b}_m$ are the learnable parameters
of the MLP projection.
Each path representation is matched only with its corresponding semantic prompt.

The semantic router captures preferences from the current graph's path representations, allowing the meta-path weights to adapt to different domains.
We combine the base and semantic scores to integrate statistical routing cues with path-specific semantic information.
The resulting routing weights are:
\begin{equation}
\alpha_{v,p}^{m}
=
\frac{
\exp\!\left(s_{v,p}^{m,\mathrm{base}}
+\gamma_{\mathrm{prompt}}s_{v,p}^{m,\mathrm{prompt}}\right)
}{
\sum_{j\in\mathcal{P}_m}
\exp\!\left(s_{v,j}^{m,\mathrm{base}}
+\gamma_{\mathrm{prompt}}s_{v,j}^{m,\mathrm{prompt}}\right)
},
\label{eq:final_router}
\end{equation}
where $\gamma_{\mathrm{prompt}}$ controls the prompt contribution.
The routed CLS representation for modality $m$ is:
\begin{equation}
\mathbf{Z}_v^m
=
\sum_{p\in\mathcal{P}_m}
\alpha_{v,p}^{m}\mathbf{h}_{v,p}^{m}.
\label{eq:metapath_aggregation}
\end{equation}
Text and image paths are aggregated independently, reducing the path-specific representations to CLS tokens $\mathbf{Z}_v^T$ and $\mathbf{Z}_v^I$.

After this aggregation, we reinsert the routed CLS representations into the corresponding complete token sequences:
\begin{equation}
\mathbf{H}_{m,v}^{\mathrm{next}}
=
\left[
(\mathbf{Z}_v^m)^\top;
\widetilde{\mathbf{H}}_{m,v}^{\mathrm{body}}
\right],
\qquad m\in\{T,I\},
\label{eq:cls_reinsertion}
\end{equation}
where $m=T$ and $m=I$ correspond to the text and image sequences,
respectively, $\widetilde{\mathbf{H}}_{m,v}^{\mathrm{body}}$ contains the contextualized tokens excluding CLS, and $[\,;\,]$ denotes concatenation along the sequence dimension.
The body tokens remain unchanged during MGP and MSPR.
The next IMC block processes these updated sequences, allowing the
routed neighborhood information to interact with local token content.
Repeating this process alternates IMC with the combined MGP--MSPR block.
The final outputs are two modality-specific sequences
$\mathbf{H}_{txt,v}^{\mathrm{out}}$ and
$\mathbf{H}_{img,v}^{\mathrm{out}}$, whose first tokens are denoted by
$\mathbf{h}_{txt,v}^{\mathrm{CLS,out}}$ and
$\mathbf{h}_{img,v}^{\mathrm{CLS,out}}$, respectively.

\subsubsection{Contextual Token Compression with Q-Former}
To reduce the redundant information in the tokens sequence and improve the LLM inference efficiency, we compress the body tokens of
each final modality sequence using a Q-Former~\cite{li2023blip}.
For $m\in\{T,I\}$, let
$\mathbf{Q}_m\in\mathbb{R}^{n_q\times d}$ contain $n_q$ learnable
queries. The compressed tokens are
\begin{equation}
\mathbf{Q}_v^m
=
\operatorname{CrossAttn}_m
\left(
\mathbf{Q}_m,
\mathbf{H}_{m,v}^{\mathrm{out,body}},
\mathbf{H}_{m,v}^{\mathrm{out,body}}
\right),
\label{eq:qformer_compression}
\end{equation}
where $\mathbf{H}_{m,v}^{\mathrm{out,body}}$ excludes the final CLS
token, and the three arguments denote queries, keys, and values.
The CLS token is retained without compression, yielding the soft-token
sequence
$\mathbf{S}_v^m=[(\mathbf{h}_{m,v}^{\mathrm{CLS,out}})^\top;
\mathbf{Q}_v^m]\in\mathbb{R}^{(n_q+1)\times d}$
for subsequent instruction tuning.

\subsubsection{Self-supervised Pre-training}
\label{sec:self_supervised_pretraining}
We jointly optimize the HMP encoder and Q-Former with self-supervised objectives that preserve modality semantics and graph structures while retaining informative content during token compression.

\noindent\textbf{HMP Reconstruction.}
We optimize HMP using feature and structure reconstruction losses to learn CLS representations in a unified feature space.
For feature reconstruction, we randomly mask the CLS representations of a subset of nodes.
For modality $m\in\{T,I\}$, the masked CLS token is replaced by a modality-specific learnable mask vector shared across all source domains:
\begin{equation}
\widetilde{\mathbf{h}}_{v}^{m,\mathrm{CLS}}
=
\begin{cases}
\mathbf{e}_{\mathrm{mask}}^{m},
& v\in\mathcal{V}_{\mathrm{mask}},\\
\mathbf{h}_{v}^{m,\mathrm{CLS}},
& \mathrm{otherwise},
\end{cases}
\label{eq:cls_mask}
\end{equation}
where $\mathbf{h}_{v}^{m,\mathrm{CLS}}$ is the original CLS representation
of node $v$, $\mathbf{e}_{\mathrm{mask}}^{m}$ is the learnable mask vector,
and $\mathcal{V}_{\mathrm{mask}}$ denotes the masked node set.

After meta-path propagation and routing, each source domain $d\in\mathcal{D}_s$
and modality $m$ employs an independent decoder $D_d^m(\cdot)$ to reconstruct
the original CLS representation:
\begin{equation}
\begin{aligned}
\widehat{\mathbf{h}}_{v}^{m,\mathrm{CLS}}
&=
D_d^m(\mathbf{Z}_v^m),\\
\mathcal{L}_{\mathrm{feat}}
&=
\sum_{d\in\mathcal{D}_s}
\frac{1}{|\mathcal{V}_{\mathrm{mask}}^d|}
\sum_{v\in\mathcal{V}_{\mathrm{mask}}^d}
\sum_{m\in\{T,I\}}
\left\|
\widehat{\mathbf{h}}_{v}^{m,\mathrm{CLS}}
-
\mathbf{h}_{v}^{m,\mathrm{CLS}}
\right\|_2^2 ,
\end{aligned}
\label{eq:feature_reconstruction}
\end{equation}
where $\mathbf{Z}_v^m$ is the routed HMP representation,
$\mathcal{D}_s$ is the set of source domains, and
$\mathcal{V}_{\mathrm{mask}}^d$ denotes the masked nodes in domain $d$.

We preserve graph topology by predicting edge probabilities
$\widehat{a}_{vu}^{m}
=\sigma(g_m(\mathbf{Z}_v^m,\mathbf{Z}_u^m))$
and minimizing
\begin{equation}
\mathcal{L}_{\mathrm{edge}}
=
\sum_{\substack{d\in\mathcal{D}_s\\m\in\{T,I\}}}
\sum_{y\in\{0,1\}}
\mathbb{E}_{(v,u)\sim\mathcal{E}_d^{y}}
\left[
\operatorname{BCE}(\widehat{a}_{vu}^{m},y)
\right],
\label{eq:structure_reconstruction}
\end{equation}
where $g_m$ is the edge decoder and $\sigma$ is the sigmoid function.
Here, $\mathcal{E}_d^{1}=\mathcal{E}_d^{+}$ and
$\mathcal{E}_d^{0}=\mathcal{E}_d^{-}$ denote observed and sampled
negative edges in source domain $d$, respectively.
Each expectation averages uniformly over the corresponding edge set,
and $\operatorname{BCE}(a,y)=-y\log a-(1-y)\log(1-a)$.

\begin{table*}[t]
\centering
\caption{Statistics and characteristics of the multimodal graph datasets.
Node and edge counts follow the original benchmark reports before
modality-based node splitting. \#Classes denotes the number of node
classes; ``--'' indicates that node classification labels are unavailable.}
\label{tab:dataset_statistics}
\footnotesize
\setlength{\tabcolsep}{3pt}
\renewcommand{\arraystretch}{1.08}
\begin{adjustbox}{max width=\textwidth}
\begin{tabular}{@{}llrrclll@{}}
\toprule
\textbf{Dataset}
& \textbf{Domain (Source)}
& \textbf{\#Nodes}
& \textbf{\#Edges}
& \textbf{\#Classes}
& \textbf{Text Attributes}
& \textbf{Image Attributes}
& \textbf{Edge Semantics} \\
\midrule

Toys
& E-commerce (Amazon)
& 20,695
& 126,886
& 18
& Titles and descriptions
& Product images
& Co-purchase/co-click \\

Reddit-S
& Social media (Reddit)
& 15,894
& 566,160
& 20
& Post text
& Post images
& Shared commenters \\

Amazon-Sports
& E-commerce (Amazon)
& 50,250
& 356,202
& --
& Product titles
& Product images
& Co-purchase \\

Movies
& E-commerce (Amazon)
& 16,672
& 218,390
& 20
& Titles and descriptions
& Product images
& Co-purchase/co-click \\

Grocery
& E-commerce (Amazon)
& 17,074
& 171,340
& 20
& Titles and descriptions
& Product images
& Co-purchase/co-click \\

Ele-fashion
& E-commerce (Amazon)
& 97,766
& 199,602
& 12
& Fashion titles
& Fashion images
& Co-purchase \\

Goodreads
& Books (Goodreads)
& 685,294
& 7,235,084
& 10
& Book descriptions
& Book covers
& User-interest associations \\

\bottomrule
\end{tabular}
\end{adjustbox}
\vspace{-0.0cm}
\end{table*}

\noindent\textbf{Q-Former Compression.}
To preserve modality semantics during compression, we align the
original body-token and compressed query representations using
symmetric InfoNCE. Defining
$s_{ij}^m=(\mathbf{z}_{v_i,\mathrm{stu}}^m)^\top
\mathbf{z}_{v_j,\mathrm{tea}}^m/\tau$, the loss for node $v_i$ is
\begin{equation}
\mathcal{L}_{\mathrm{global}}(v_i)
=
-\frac{1}{2}
\sum_{m\in\{T,I\}}
\log
\frac{
e^{2s_{ii}^m}
}{
\left(\sum_j e^{s_{ij}^m}\right)
\left(\sum_j e^{s_{ji}^m}\right)
},
\label{eq:global_contrastive}
\end{equation}
where $\mathbf{z}_{v_i,\mathrm{tea}}^m$ and
$\mathbf{z}_{v_i,\mathrm{stu}}^m$ are obtained by mean-pooling
the original body tokens $\mathbf{H}_{m,v_i}^{\mathrm{body}}$
and Q-Former outputs $\mathbf{Q}_{v_i}^m$, respectively,
and $\tau$ is the contrastive temperature.
The sums over $j$ include the matching node $v_i$ and comparison
nodes: same-node representations form positive pairs, while
different-node representations form negative pairs.
The two denominator terms correspond to student-to-teacher and
teacher-to-student matching, respectively.

Finally, the self-supervised pre-training objective is
\begin{equation}
\mathcal{L}_{\mathrm{SSL}}
=
\lambda_{\mathrm{feat}}\mathcal{L}_{\mathrm{feat}}
+
\lambda_{\mathrm{edge}}\mathcal{L}_{\mathrm{edge}}
+
\lambda_{\mathrm{global}}\mathcal{L}_{\mathrm{global}},
\label{eq:ssl_total}
\end{equation}
where $\lambda_{\mathrm{feat}}$, $\lambda_{\mathrm{edge}}$, and
$\lambda_{\mathrm{global}}$ weight the feature reconstruction,
structure reconstruction, and global contrastive objectives,
respectively.

\subsection{Evidence-Guided Instruction Tuning}
\label{sec:evidence_instruction_tuning}
Relevant neighboring evidence enriches the query's multimodal context.
Each evidence node adaptively integrates textual and visual information from its own neighborhood and its graph-aware soft tokens provide contextual cues to support LLM-based prediction.

\noindent\textbf{Evidence Selection.}
We select evidence from the training split and denote the evidence
set for query $q$ by $\mathcal{E}_q$.
For node classification, $q=v$ is a target node and each evidence
item $e=u$ is a neighboring node with class label $y_e=y_u$.
For link prediction, $q=(v,w)$ is a target node pair and each
evidence item $e=(a,b)$ is a node pair with binary edge label
$y_e\in\{0,1\}$.
Let $\mathcal{V}(z)$ denote the nodes contained in a query or
evidence item $z$: $\mathcal{V}(v)=\{v\}$ and
$\mathcal{V}((v,w))=\{v,w\}$.

Firstly, we score each neighboring node $u$
by averaging its modality-wise CLS similarities with the
target node $v$:
\begin{equation}
s(v,u)
=
\frac{1}{2}\sum_m
\operatorname{cos}\!\left(
\mathbf{h}_{m,v}^{\mathrm{CLS,out}},
\mathbf{h}_{m,u}^{\mathrm{CLS,out}}
\right),
\label{eq:evidence_similarity}
\end{equation}
where $m\in\{\mathrm{txt},\mathrm{img}\}$ denotes the modality,
$\mathbf{h}_{m,v}^{\mathrm{CLS,out}}$ is the final graph-aware CLS
representation of node $v$ in modality $m$, and
$\operatorname{cos}(\cdot,\cdot)$ denotes cosine similarity.

For query $q=v$, the evidence set is selected as
$\mathcal{E}_q=\operatorname{TopK}_{u\in\mathcal{C}(v)}s(v,u)$,
where $\mathcal{C}(v)$ denotes the candidate neighborhood of $v$
restricted to the training split.
We primarily consider one-hop neighbors and additionally include
two-hop neighbors when insufficient candidates are available,
retaining up to $K$ evidence nodes.

For link prediction, we use positive and negative node pairs
as evidence, with binary labels indicating whether their
endpoints are connected, rather than node-class labels.
The positive-to-negative evidence ratio is set to $1:3$.
Given a query pair $(v,w)$, positive candidates are observed
edges in the subgraph induced by the endpoints' combined
two-hop neighborhoods, excluding the query edge.
Using the node similarity in Eq.~\ref{eq:evidence_similarity},
we score a candidate pair $(a,b)$ as
\begin{equation}
s_{\mathrm{LP}}\big((v,w),(a,b)\big)
=
\frac{1}{2}\max\left\{
\begin{aligned}
&s(v,a)+s(w,b),\\
&s(v,b)+s(w,a)
\end{aligned}
\right\},
\label{eq:lp_evidence_similarity}
\end{equation}
which accounts for both endpoint alignments.
Negative candidates are constructed by replacing one endpoint
of a retrieved positive edge with a randomly sampled node,
excluding self-pairs, known positive edges, and reserved
query pairs.
For query $q=(v,w)$, we rank positive and negative candidates
separately using $s_{\mathrm{LP}}$.
We select the highest-scoring pairs at the prescribed ratio
to form $\mathcal{E}_q$, with up to $K$ pairs in total.

\noindent\textbf{Soft-token Projection.}
Let $\mathbf{S}_x^T$ and $\mathbf{S}_x^I$ denote the complete text and
image soft-token sequences of node $x$, respectively. We project every
soft token into the LLM embedding space using modality-specific MLP
projectors:
\begin{equation}
\widetilde{\mathbf{S}}_x^m
=
P_{\phi}^{m}
\left(
\mathbf{S}_x^m
\right),
\qquad
m\in\{T,I\},
\quad
x\in\{v\}\cup\mathcal{E}_v,
\label{eq:soft_token_projection}
\end{equation}
where $P_{\phi}^{m}(\cdot)$ denotes the learnable projector for modality
$m$. The projection is applied token-wise and preserves the complete
soft-token sequence rather than pooling it into a single vector.

\begin{table*}[t]
\centering
\caption{Overall performance comparison on node classification and link prediction.
Results are reported in percentages.
Bold and underlining indicate the best and second-best results, respectively.
}
\label{tab:main_results}
\small
\setlength{\tabcolsep}{3pt}
\renewcommand{\arraystretch}{1.08}
\begin{tabular}{
@{}c@{\hspace{4pt}}|@{\hspace{4pt}}
>{\raggedright\arraybackslash}m{2.8cm}
@{\hspace{3pt}}|@{\hspace{3pt}}*{8}{c}@{}
}
\toprule
\multirow[c]{2}{*}{\textbf{Metric}}
& \multicolumn{1}{c@{\hspace{3pt}}|@{\hspace{3pt}}}{\multirow[c]{2}{*}{\textbf{Method}}}
& \multicolumn{4}{c}{\textbf{Node Classification}}
& \multicolumn{4}{c}{\textbf{Link Prediction}} \\
\cline{3-6}\cline{7-10}
& & \textbf{Movies} & \textbf{Grocery} & \textbf{Ele-fashion} & \textbf{Goodreads}
& \textbf{Movies} & \textbf{Grocery} & \textbf{Ele-fashion} & \textbf{Goodreads} \\
\midrule

\multirow{9}{*}{\rotatebox[origin=c]{90}{\textbf{Accuracy (\%)}}}
& Qwen2.5-VL-7B
& 17.86 & 67.06 & 64.69 & 35.27
& \underline{96.14} & \underline{95.89} & \underline{96.08} & \textbf{95.25} \\

& Vicuna-7B-v1.5
& 10.81 & 39.42 & 44.73 & 27.11
& 4.76 & 4.76 & 8.23 & 6.78 \\

& LLaVA-1.5-7B
& 19.02 & 39.48 & 43.91 & 38.56
& 56.11 & 58.25 & 89.79 & 33.69 \\

& GraphPrompter
& 18.04 & 52.33 & 37.85 & 38.66
& 95.81 & 95.72 & 94.18 & \underline{95.17} \\

& GraphGPT
& 9.82 & 40.38 & 30.87 & 25.12
& 29.07 & 40.73 & 18.10 & 36.87 \\

& LLaGA-ND
& 3.31 & 9.39 & 8.80 & 10.24
& 49.67 & 92.64 & 75.99 & 27.55 \\

& Mario
& 24.87 & 52.59 & 51.88 & 57.12
& 95.18 & \textbf{96.07} & \textbf{96.54} & 94.92 \\

& MLaGA
& \underline{33.15} & \underline{76.91} & \underline{68.90} & \underline{63.07}
& 85.77 & 82.34 & 94.80 & 86.06 \\

& \textbf{MAGE}
& \textbf{35.71} & \textbf{78.10} & \textbf{75.02} & \textbf{72.40}
& \textbf{96.38} & 95.18 & 95.09 & 94.33 \\
\midrule

\multirow{9}{*}{\rotatebox[origin=c]{90}{\textbf{Macro-F1 (\%)}}}
& Qwen2.5-VL-7B
& 15.09 & 60.47 & 41.56 & 30.62
& \underline{65.76} & \underline{69.51} & 64.08 & 48.98 \\

& Vicuna-7B-v1.5
& 5.00 & 35.55 & 17.36 & 19.04
& 4.55 & 4.55 & 8.21 & 6.71 \\

& LLaVA-1.5-7B
& 7.52 & 35.05 & 14.04 & 33.51
& 41.84 & 43.17 & 69.81 & 29.19 \\

& GraphPrompter
& 11.29 & 51.36 & 26.61 & 35.97
& 60.33 & 65.03 & 77.72 & 51.38 \\

& GraphGPT
& 6.15 & 35.95 & 19.13 & 23.16
& 25.83 & 33.05 & 18.36 & 30.20 \\

& LLaGA-ND
& 0.89 & 0.86 & 1.68 & 1.72
& 38.74 & 52.66 & 46.98 & 24.41 \\

& Mario
& 24.22 & 55.02 & 35.90 & 49.83
& 62.61 & 66.47 & 75.25 & 51.53 \\

& MLaGA
& \underline{32.64} & \underline{67.67} & \underline{48.55} & \underline{60.24}
& 64.15 & 61.35 & \underline{77.74} & \underline{60.45} \\

& \textbf{MAGE}
& \textbf{33.47} & \textbf{71.58} & \textbf{52.31} & \textbf{63.99}
& \textbf{81.93} & \textbf{78.45} & \textbf{78.66} & \textbf{65.93} \\
\bottomrule
\end{tabular}
\end{table*}

\noindent\textbf{Evidence-Enhanced Instruction.}
We denote the complete projected soft-token sequence of an
evidence item $e$ by $\widetilde{\mathbf{S}}_e$:
\[
\widetilde{\mathbf{S}}_e
=
\begin{cases}
\left[
\widetilde{\mathbf{S}}_u^T,\,
\widetilde{\mathbf{S}}_u^I
\right],
& e=u,\\[2pt]
\left[
\widetilde{\mathbf{S}}_a^T,\,
\widetilde{\mathbf{S}}_a^I,\,
\widetilde{\mathbf{S}}_b^T,\,
\widetilde{\mathbf{S}}_b^I
\right],
& e=(a,b).
\end{cases}
\]
Here, brackets denote sequence concatenation, preserving
the endpoint order for node-pair evidence.

For query $q$, we combine the task instruction $\mathcal{P}_q$,
the raw content and projected soft tokens of all nodes
$x\in\mathcal{V}(q)$, and the representations and labels
of all evidence items $e\in\mathcal{E}_q$:
\begin{equation}
\mathbf{C}_q=\Big[
\mathcal{P}_q,\,
\{(\mathcal{T}_x,\mathcal{I}_x,
\widetilde{\mathbf{S}}_x^T,\widetilde{\mathbf{S}}_x^I)\},\,
\{(\widetilde{\mathbf{S}}_e,y_e)\}
\Big].
\label{eq:evidence_instruction}
\end{equation}

Given the target output sequence
$Y_q=(w_1,\ldots,w_{L_q})$, instruction tuning is performed
using the autoregressive objective
\begin{equation}
\mathcal{L}_{\mathrm{IT}}
=
-
\sum_{q\in\mathcal{D}_{\mathrm{train}}}
\sum_{t=1}^{L_q}
\log
P_{\theta}
\left(
w_t
\mid
\mathbf{C}_q,
w_{<t}
\right),
\label{eq:instruction_tuning}
\end{equation}
where $\mathcal{D}_{\mathrm{train}}$ contains source-domain
training queries from both node classification and link prediction,
$\theta$ denotes the frozen LLM parameters, and $\phi$ denotes
the trainable projector parameters.
Gradients are propagated through the frozen LLM to optimize
the projectors while keeping the LLM unchanged.

\section{Experiments}
\label{sec:experiments}
We conduct extensive experiments to evaluate MAGE for multimodal
graph prediction on unseen target datasets. Specifically, we
investigate the following five research questions:
\textbf{RQ1:} How does MAGE compare with existing baselines on node classification and link prediction across unseen target datasets?
\textbf{RQ2:} How does each key component contribute to MAGE's predictive performance?
\textbf{RQ3:} How sensitive is MAGE to the evidence budget and the number of Q-Former query tokens?
\textbf{RQ4:} How do token and evidence budgets affect MAGE's runtime, and how does its preparation cost scale with graph size?
\textbf{RQ5:} How do meta-path routing preferences vary across datasets, nodes, and terminal modalities?

\subsection{Experiment Setup}
\noindent\textbf{Datasets.}
\label{sec:datasets}
We conduct experiments on seven multimodal graph datasets spanning diverse
domains. \textbf{Toys}, \textbf{Amazon-Sports}, \textbf{Movies}, and
\textbf{Grocery} are e-commerce product graphs from Amazon~\cite{ni2019justifying,yan2024graph}, covering
different product categories. \textbf{Reddit-S} is a social-media graph
constructed from Reddit~\cite{desai2021redcaps,yan2024graph}. \textbf{Ele-fashion} is an e-commerce graph of fashion products~\cite{ni2019justifying,zhu2025mosaic}, while \textbf{Goodreads} is a book-domain graph for book recommendation~\cite{wan2018item,zhu2025mosaic}. 
Some high-level summaries of these datasets are shown in Table~\ref{tab:dataset_statistics}. 

\noindent\textbf{Baselines.}
We compare MAGE with several categories of baselines. The first category is general language and multimodal language models, including Vicuna-7B-v1.5~\cite{zheng2023judging}, Qwen2.5-VL-7B~\cite{bai2025qwen25vl}, and LLaVA-1.5-7B~\cite{liu2024improved}. The second category is graph-LLM methods, including GraphPrompter~\cite{liu2024can}, GraphGPT~\cite{tang2024graphgpt}, and LLaGA-ND~\cite{chen2024llaga}. The last category is multimodal graph-LLM methods: Mario~\cite{sun2026mario} and MLaGA~\cite{fan2026mlaga}.
For each baseline, we use the LLM backbone and hyperparameter settings
reported in its original paper.
Mario and MLaGA also use neighbor labels, following their original
experimental protocols.

\noindent\textbf{Implementation Details.}
\label{sec:implementation}
We adopt a multi-source transfer setting, where Toys, Reddit-S, and Amazon-Sports are jointly used for training, while Movies, Grocery, Ele-fashion, and Goodreads are used as unseen target datasets for evaluation.
All datasets are partitioned into training, validation, and test splits following the protocols in their respective source papers. For each target dataset, evaluation is performed on the test split, evidence is drawn exclusively from the training split, and no model parameters are updated.

We employ pretrained T5~\cite{raffel2020exploring} and ViT~\cite{dosovitskiy2020image} encoders to encode textual and visual content, respectively, and use Qwen2.5-VL-7B-Instruct as the LLM backbone. MAGE contains two HMP layers,
which are alternately stacked with Transformer blocks. Each modality uses 16 Q-Former queries to compress its token sequence, and the CLS masking ratio during self-supervised pre-training is set to $0.3$. For evidence construction, we select 8 neighboring nodes for node classification and 12 neighboring node pairs for link prediction.

MAGE is trained in two stages. In Stage 1, the graph representation modules are jointly pretrained on Toys, Reddit-S, and Amazon-Sports using the self-supervised objective in Eq.~\eqref{eq:ssl_total}. In Stage 2, the shared MLP projectors are jointly optimized using node-classification instructions from Toys and Reddit-S and link-prediction instructions from Amazon-Sports, while the pretrained graph representation modules and the LLM remain frozen.

For both node classification and link prediction, we report Accuracy and Macro-F1. 
All positive test edges are held out from the graph before preprocessing and representation generation.
Link prediction is formulated as a binary classification task that predicts whether an edge exists between a given pair of nodes.
For each dataset, we construct the evaluation set with 500 positive edges and 10,000 sampled negative edges.

\begin{table*}[t]
\centering
\caption{Ablation study on node classification and link prediction.
Results are reported as Macro-F1 (\%), with the best result in each column
highlighted in bold. ``--'' indicates unavailable results.}
\label{tab:ablation_f1}
\setlength{\tabcolsep}{5pt}
\renewcommand{\arraystretch}{1.12}
\resizebox{\textwidth}{!}{%
\begin{tabular}{lcccccccc}
\toprule
& \multicolumn{4}{c}{\textbf{Node Classification}}
& \multicolumn{4}{c}{\textbf{Link Prediction}} \\
\cmidrule(lr){2-5}
\cmidrule(lr){6-9}
\textbf{Variant}
& \textbf{Movies}
& \textbf{Grocery}
& \textbf{Ele-fashion}
& \textbf{Goodreads}
& \textbf{Movies}
& \textbf{Grocery}
& \textbf{Ele-fashion}
& \textbf{Goodreads} \\
\midrule

w/o Soft Tokens
& 16.47 & 62.02 & 42.80 & 44.68
& 53.75 & 59.36 & 51.72 & 49.94 \\

w/o Meta-paths
& 27.42 & 67.71 & 47.49 & 53.68
& 56.59 & 64.70 & 68.23 & 50.41 \\

Uniform Meta-path Fusion
& 28.45 & 70.26 & 50.54 & 57.50
& 80.08 & 77.79 & 77.28 & 64.96 \\

w/o Evidence ($K_e=0$)
& 10.01 & 48.04 & 37.79 & 30.34
& 76.91 & 74.76 & 77.76 & 59.41 \\

random Evidence
& 32.86 & 70.64 & 49.75 & 59.13
& 74.22 & 72.35 & 61.96 & 57.08 \\

w/o Reconstruction
& 32.51 & 69.52 & 50.68 & 60.79
& 79.34 & 71.78 & 78.34 & 58.78 \\

w/o Q-Former all
& 28.98 & 68.69 & 50.88 & 55.20
& 80.28 & 72.97 & 76.71 & 58.33 \\

w/o Q-Former compression
& 29.23 & 70.69 & 48.95 & 56.68
& 66.58 & 52.69 & 78.45 & 57.08 \\

\midrule
\textbf{MAGE}
& \textbf{33.47} 
& \textbf{71.58}
& \textbf{52.31} 
& \textbf{63.99}
& \textbf{81.93} 
& \textbf{78.45}
& \textbf{78.66} 
& \textbf{65.93} \\

\bottomrule
\end{tabular}%
}
\vspace{2pt}
\begin{minipage}{\textwidth}
\footnotesize
\end{minipage}
\vspace{-0.5cm}
\end{table*}

\subsection{Overall Performance (RQ1).}
Table~\ref{tab:main_results} compares MAGE with the baselines on four
target datasets for node classification and link prediction.
MAGE achieves the highest Macro-F1 in all eight task--dataset
combinations and the best accuracy on all four node classification
datasets.

For node classification, MAGE improves accuracy over the strongest
baseline on Movies, Grocery, Ele-fashion, and Goodreads by 2.56, 1.19,
6.12, and 9.33 percentage points, respectively. Its corresponding
Macro-F1 improvements are 0.83, 3.91, 3.76, and 3.75 points.
Averaged equally across the four datasets, MAGE achieves 65.31\%
accuracy and 55.34\% Macro-F1, outperforming MLaGA by 4.80 and
3.06 points, respectively. These results demonstrate consistent
improvements across the evaluated target graphs rather than gains
restricted to a single dataset.

For link prediction, MAGE exceeds the strongest baseline in Macro-F1
by 16.17, 8.94, 0.94, and 5.48 points on Movies, Grocery, Ele-fashion,
and Goodreads, respectively. Although MAGE does not achieve the
highest accuracy on every dataset, accuracy alone is less informative
under the evaluation ratio of 500 positive to 10,000 negative edges,
where predicting every pair as a non-edge already yields 95.24\%
accuracy. The consistent Macro-F1 improvements therefore provide
an important complementary assessment of its link prediction
performance. Overall, these results support the effectiveness of
MAGE for multimodal graph prediction under the multi-source
transfer setting.

\begin{figure}[t]
    \centering
    \includegraphics[width=\columnwidth]{
        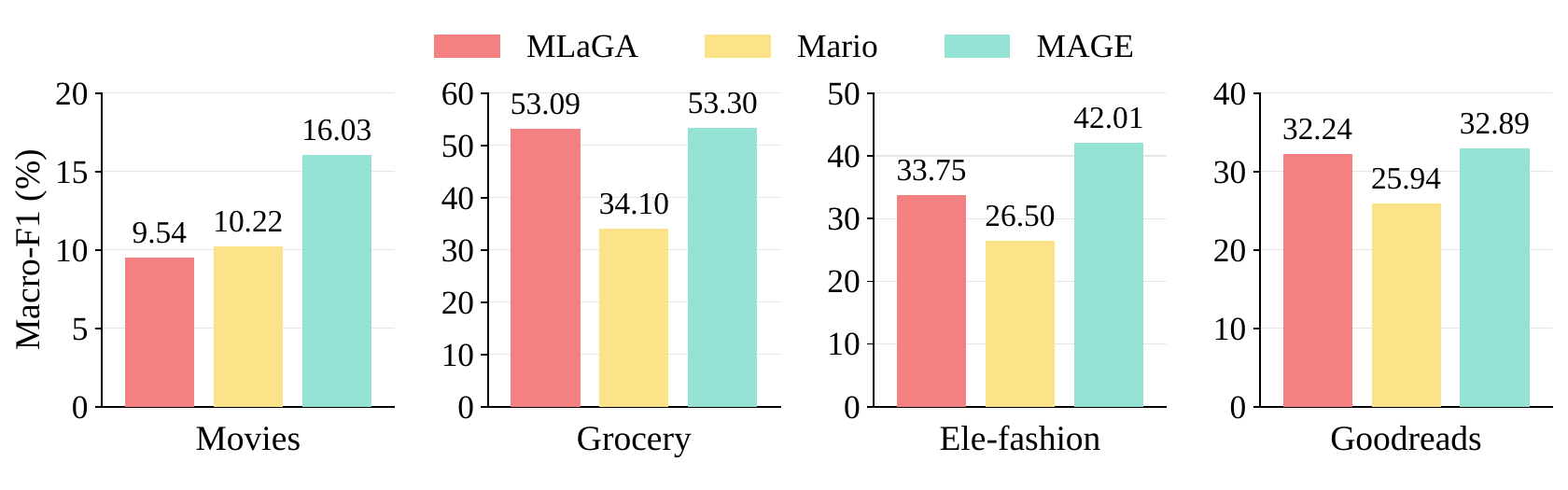
    }
    \caption{
        Zero-Shot Node Classification. 
    }
    \label{fig:nc_without_neighbor_labels}
\vspace{-0.0cm}
\end{figure}
\noindent\textit{Zero-Shot Prediction (RQ1).}
\label{zero-shot}
We evaluate zero-shot prediction on target datasets without target-domain fine-tuning or explicit evidence labels in the LLM input.
Each method retains its source-trained Stage~2 checkpoint and uses target-node information together with unlabeled neighbor soft tokens.

As shown in Fig.~\ref{fig:nc_without_neighbor_labels},
MAGE achieves higher Macro-F1 than both baselines on Movies and Ele-fashion, while performing comparably to MLaGA on Grocery and Goodreads.
These results show that MAGE retains predictive capability without explicit neighbor-label cues, although they do not isolate the contribution of neighbor representations from that of the target inputs.

\subsection{Ablation Study (RQ2)}
\label{sec:ablation}
We evaluate eight ablation variants to examine the contributions
of different components in MAGE.

\noindent\textbf{Graph-aware Representation Learning.}
\textbf{w/o Soft Tokens} removes the encoder and replaces the
LLM's soft-token inputs with raw text and images.
Its lower performance supports the use of learned graph-aware
representations to provide context for LLM prediction.
\textbf{w/o Reconstruction} removes the feature and structure
reconstruction losses from HMP.
The resulting performance decline supports the contribution
of these objectives to representation learning.

\noindent\textbf{Meta-path Propagation and Fusion.}
\textbf{w/o Meta-paths} removes the meta-paths and propagates
text and image features independently.
Its lower performance suggests that cross-modal propagation
captures useful neighborhood interactions beyond independent
unimodal propagation.
\textbf{Uniform Meta-path Fusion} replaces MSPR with equal-weight
averaging of the meta-path representations.
This variant remains competitive on Grocery, indicating that
uniform fusion can already exploit useful information from
multiple paths.
Adaptive weighting provides a further performance improvement.

\noindent\textbf{Neighboring Evidence.}
\textbf{w/o Evidence} removes neighboring evidence from the
LLM input, retaining only the query information.
Its lower performance suggests that explicit evidence complements
the neighborhood information already encoded in the query's
graph-aware soft tokens.
\textbf{Random Evidence} replaces similarity-based evidence
selection with random sampling, keeping all other settings unchanged.
Its lower performance, particularly on LP, supports the
effectiveness of selecting relevant evidence.

\noindent\textbf{Q-Former Design.}
\textbf{w/o Q-Former all} removes the entire Q-Former module,
including token-sequence compression and the associated
contrastive consistency loss.
Its lower performance supports the effectiveness of the complete
Q-Former design.
\textbf{w/o Q-Former compression} removes token-sequence compression
while retaining the contrastive loss for consistency between
the input and output token representations.
The performance decline despite retaining this loss supports
the benefit of token compression for LLM-based graph prediction.
\begin{figure}[t]
    \centering
    \includegraphics[width=0.98\linewidth]
        {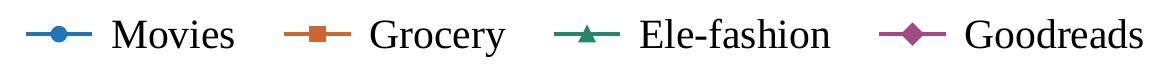}
    \par\vspace{2pt}

    \begin{subfigure}[t]{0.48\linewidth}
        \centering
        \includegraphics[width=\linewidth]
            {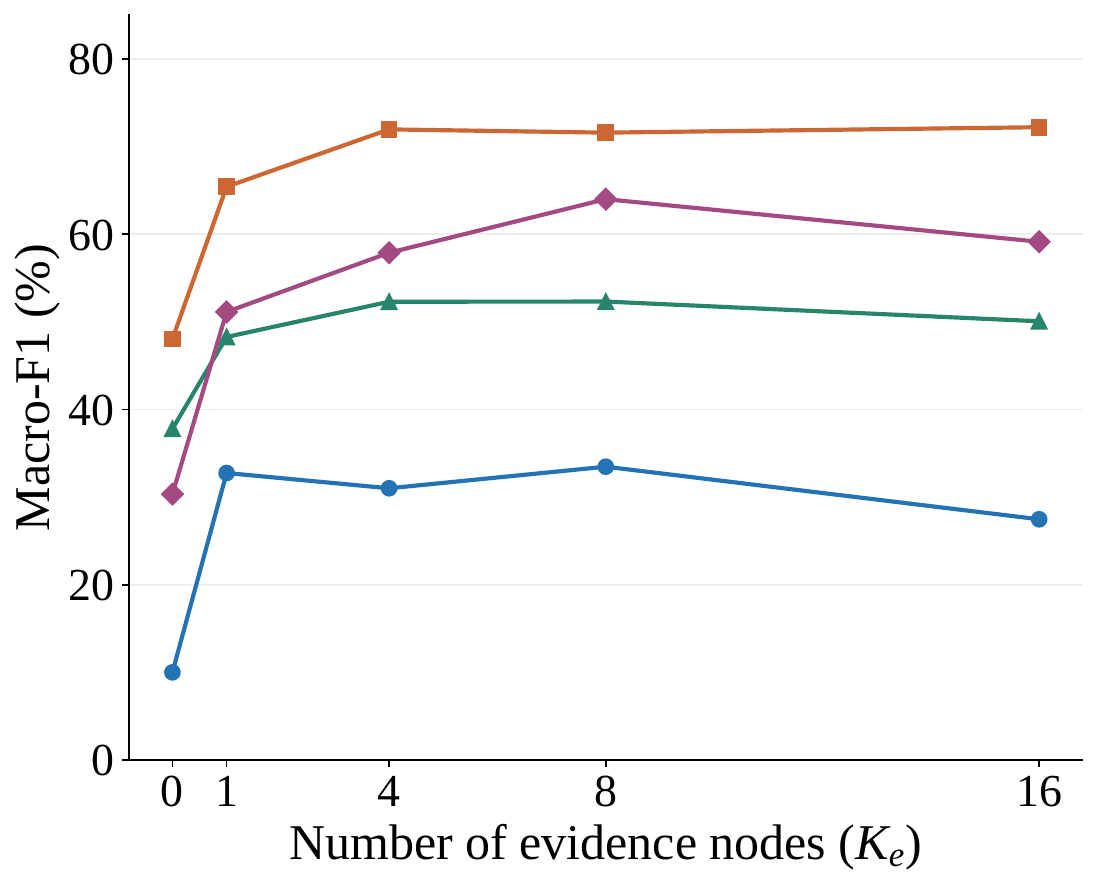}
        \caption{$K_e$: NC}
        \label{fig:ke_nc}
    \end{subfigure}\hfill
    \begin{subfigure}[t]{0.48\linewidth}
        \centering
        \includegraphics[width=\linewidth]
            {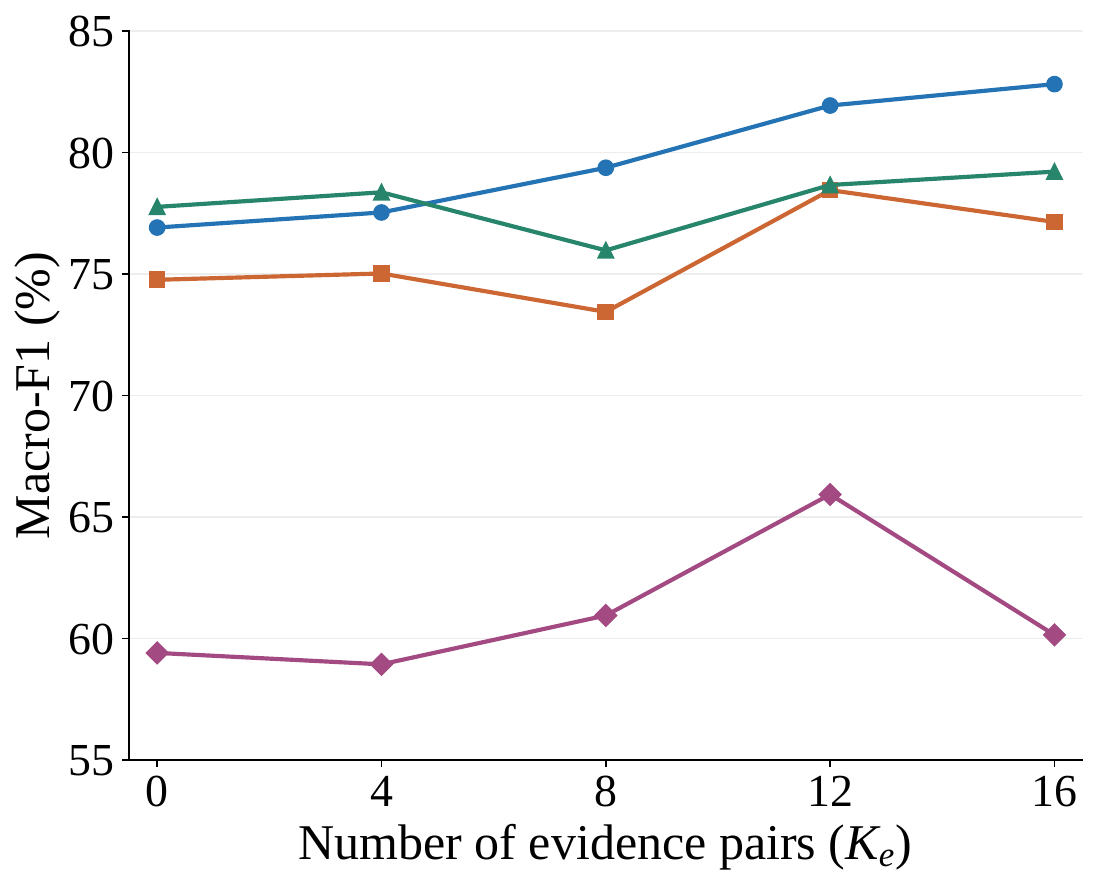}
        \caption{$K_e$: LP}
        \label{fig:ke_lp}
    \end{subfigure}

    \par\vspace{4pt}

    \begin{subfigure}[t]{0.48\linewidth}
        \centering
        \includegraphics[width=\linewidth]
            {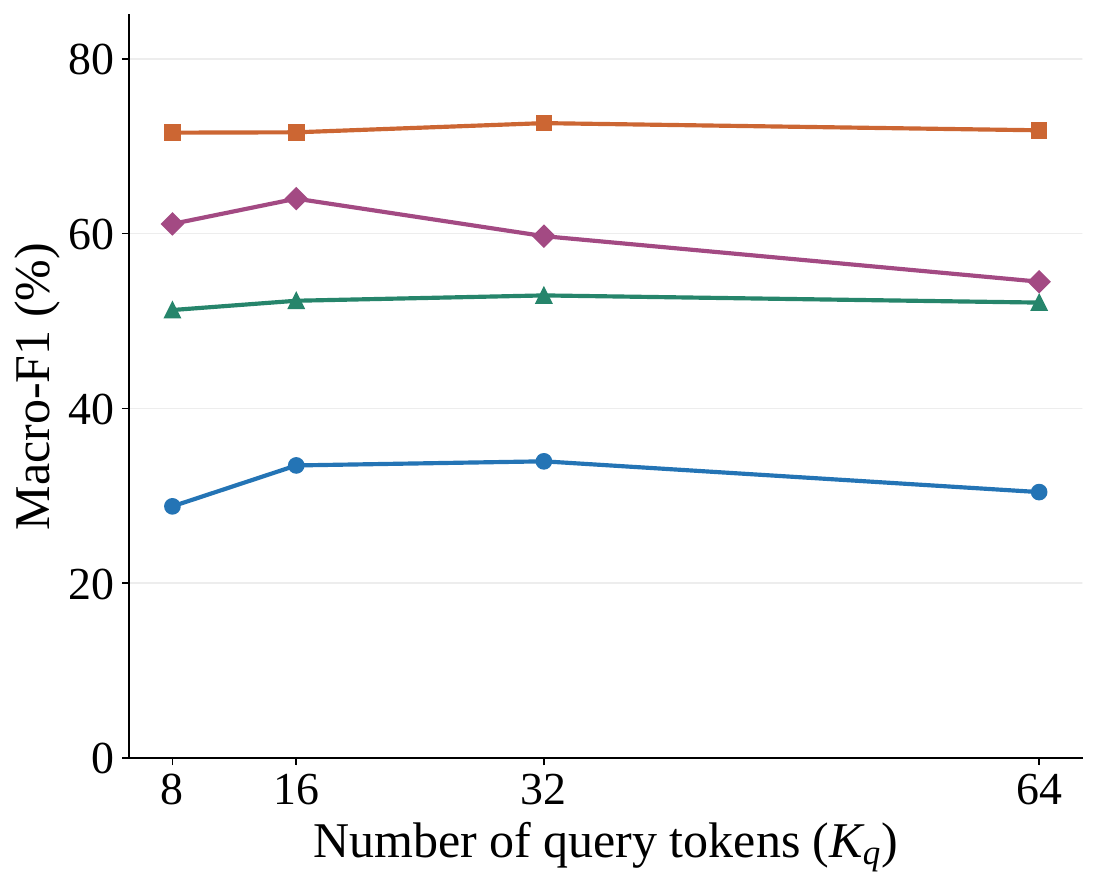}
        \caption{$K_q$: NC}
        \label{fig:kq_nc}
    \end{subfigure}\hfill
    \begin{subfigure}[t]{0.48\linewidth}
        \centering
        \includegraphics[width=\linewidth]
            {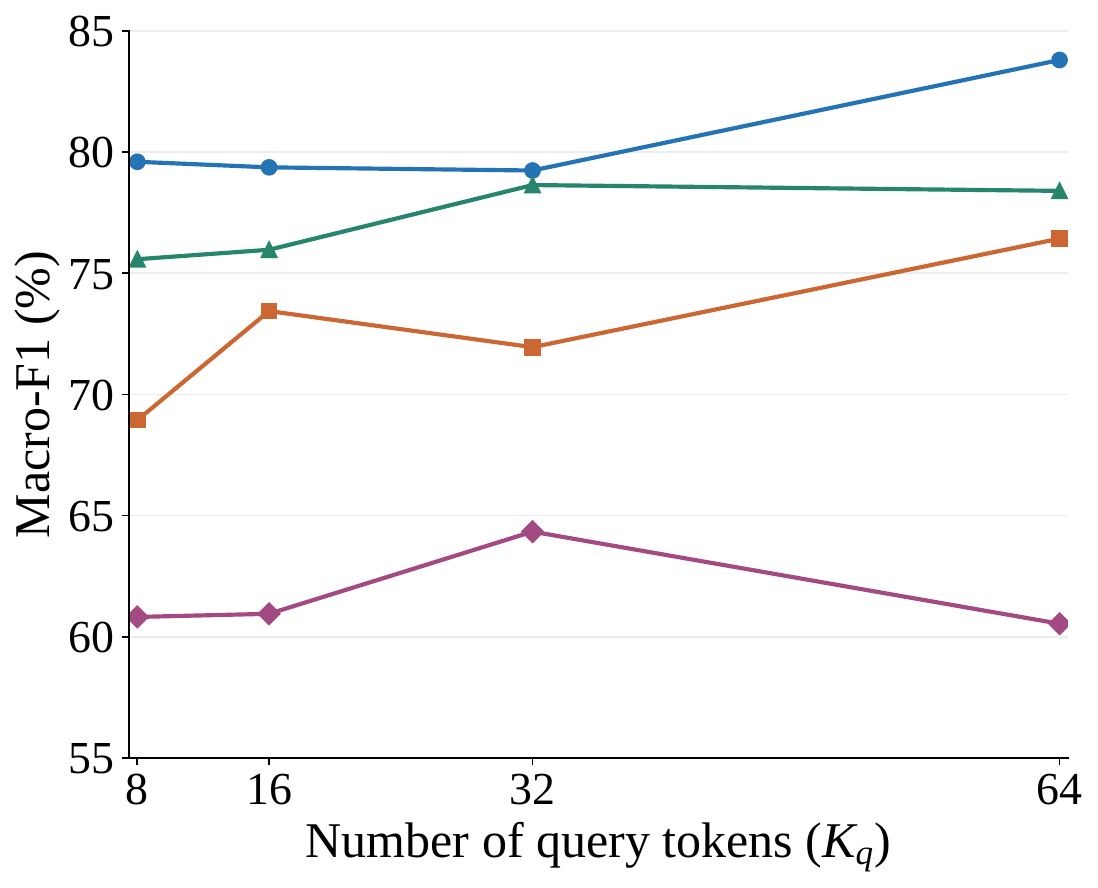}
        \caption{$K_q$: LP}
        \label{fig:kq_lp}
    \end{subfigure}

    \caption{Hyperparameter sensitivity in Macro-F1 (\%).
    We vary $K_e$ with $K_q=16$ in ~\ref{fig:ke_nc}--~\ref{fig:ke_lp},
    and vary $K_q$ with $K_e=8$ in ~\ref{fig:kq_nc}--~\ref{fig:kq_lp}.}
    \label{fig:hyperparameter_analysis}
\vspace{-0.0cm}
\end{figure}

\subsection{Hyperparameter Analysis (RQ3)}
\label{sec:hyperparameter_analysis}

We investigate the effects of the evidence budget $K_e$ and the
number of Q-Former query tokens per modality $K_q$ on Macro-F1.
Here, $K_e$ counts neighboring nodes for node classification (NC)
and neighboring node pairs for link prediction (LP).
We vary $K_e$ with $K_q=16$ and vary $K_q$ with $K_e=8$.

\noindent\textit{Effect of $K_e$.}
For NC, introducing a single evidence node substantially improves
performance over the evidence-free setting on all four datasets,
as shown in Fig.~\ref{fig:ke_nc}.
Further gains vary across datasets.
Grocery and Ele-fashion show limited improvements beyond $K_e=4$,
whereas Goodreads continues to improve up to $K_e=8$.
Movies fluctuates after the initial improvement, reaching its
highest score at $K_e=8$.
Increasing $K_e$ to 16 reduces performance on Movies,
Ele-fashion, and Goodreads compared with $K_e=8$.
These results highlight the value of neighboring evidence,
while indicating that the benefits of increasing the evidence
budget depend on the dataset.

For LP, Fig.~\ref{fig:ke_lp} shows different
responses across datasets.
Movies improves consistently as $K_e$ increases from 4 to 16.
Ele-fashion initially declines from $K_e=4$ to $8$, then
recovers and reaches its highest score at $K_e=16$.
Grocery and Goodreads attain their highest observed Macro-F1
at $K_e=12$ and decline at $K_e=16$.
These results indicate that increasing the evidence budget
does not uniformly improve link prediction, and the preferred
budget varies across datasets.

\noindent\textit{Effect of $K_q$.}
Figure~\ref{fig:kq_nc} evaluates NC performance with $K_q\in\{8,16,32,64\}$.
Movies, Grocery, and Ele-fashion achieve their highest observed
Macro-F1 at $K_q=32$, while Goodreads performs best at $K_q=16$.
Grocery and Ele-fashion remain relatively stable across the
evaluated range.
Movies improves up to $K_q=32$ before declining at $K_q=64$,
whereas Goodreads declines as $K_q$ increases beyond 16.
These observations indicate that retaining more query tokens
does not necessarily improve node classification, and the
preferred compression budget varies across datasets.

For LP, Fig.~\ref{fig:kq_lp} shows that
Movies remains relatively stable from $K_q=8$ to $32$ and
improves substantially at $K_q=64$.
Grocery exhibits a non-monotonic trend, reaching its highest
Macro-F1 at $K_q=64$.
Ele-fashion and Goodreads perform best at $K_q=32$ and decline
at $K_q=64$, with a more pronounced drop on Goodreads.
These results further indicate that increasing the query-token
budget does not uniformly improve link prediction performance.

\begin{figure}[t]
  \centering
  \includegraphics[width=0.98\linewidth]{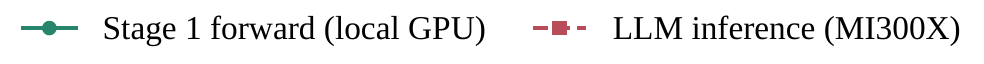}
  \par\vspace{2pt}

  \subfloat[NC: $K_q$ ($K_e=8$).
    \label{fig:movies-nc-kq}]{%
    \includegraphics[width=0.49\linewidth]{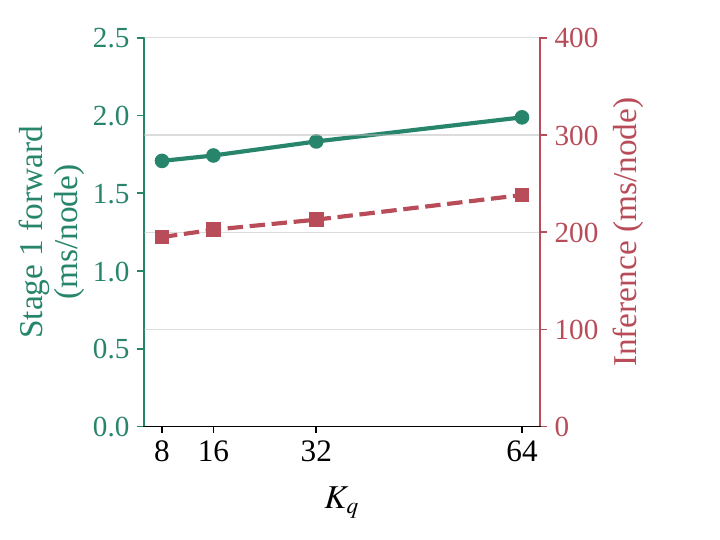}%
  }\hfill
  \subfloat[LP: $K_q$ ($K_e=8$).
    \label{fig:movies-lp-kq}]{%
    \includegraphics[width=0.49\linewidth]{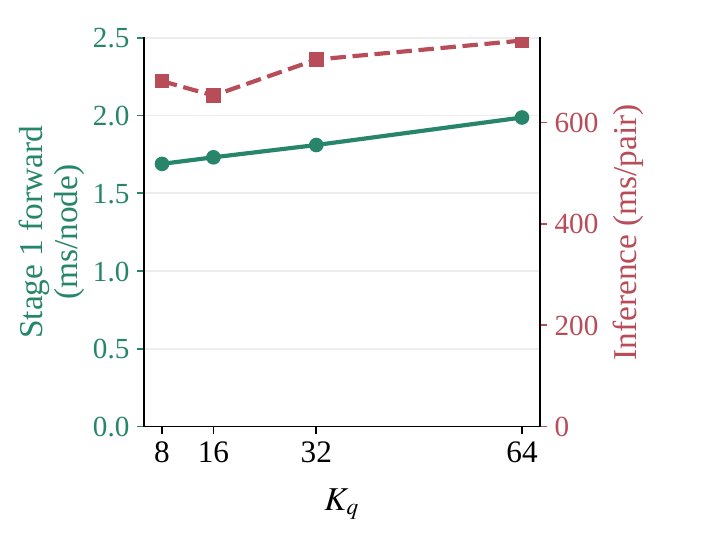}%
  }

  \par\vspace{4pt}

  \subfloat[NC: $K_e$ ($K_q=16$).
    \label{fig:movies-nc-ke}]{%
    \includegraphics[width=0.49\linewidth]{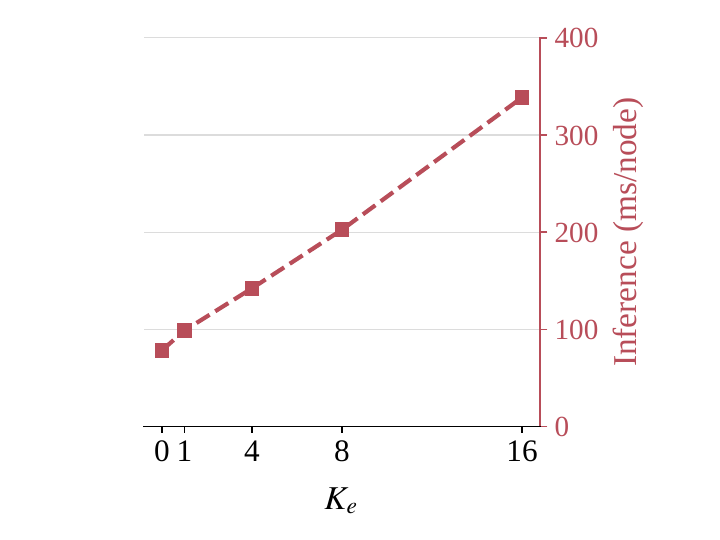}%
  }\hfill
  \subfloat[LP: $K_e$ ($K_q=16$).
    \label{fig:movies-lp-ke}]{%
    \includegraphics[width=0.49\linewidth]{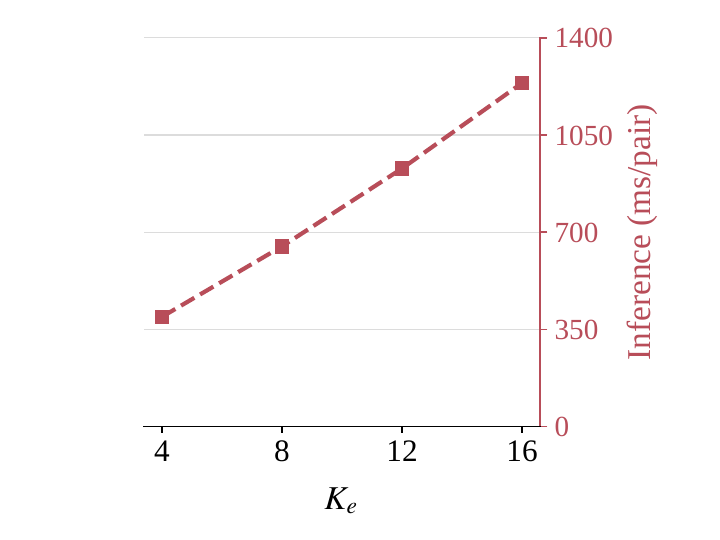}%
  }

  \caption{Amortized runtime on Movies, averaged over three runs.
  }
  \label{fig:movies-hyperparameter-timing}

\vspace{-0.5cm}
\end{figure}

\subsection{Efficiency and Scalability Analysis (RQ4)}
\noindent\textbf{Runtime under Different Token and Evidence Budgets.}
Fig.~\ref{fig:movies-hyperparameter-timing} examines the runtime
of MAGE on Movies under different query-token budgets $K_q$
and evidence budgets $K_e$.
We report Stage 1 forward time and LLM inference time separately.

\noindent\textit{Effect of $K_q$.}
With the evidence budget fixed, Stage~1 model forward time increases moderately with $K_q$ and remains similar between NC and LP, as shown in Fig.~\ref{fig:movies-nc-kq}--~\ref{fig:movies-lp-kq}.
NC inference exhibits a gradual upward trend, whereas LP inference shows a modest overall increase with non-monotonic variation across configurations.

\noindent\textit{Effect of $K_e$.}
With the query-token budget fixed, inference time increases approximately linearly with $K_e$ for both node classification and link prediction, as shown in Figs.~\ref{fig:movies-nc-ke} and~\ref{fig:movies-lp-ke}.
This trend highlights the importance of balancing predictive performance and inference efficiency when selecting the evidence budget.

LP incurs higher inference cost because each query involves two target nodes and each evidence item contains two endpoints, whereas NC uses individual neighboring nodes as evidence.
This larger input workload increases input preparation, projection, and subsequent model computation.
The measured runtime gap is dominated by input preparation and projection, rather than LLM generation alone.

\noindent\textbf{Scalability of Local Preparation.}
We examine how data preparation time changes as graph size increases. To simulate graphs of different sizes, we sample subgraphs containing 40K to 80K nodes from Ele-fashion.
We measure three stages: router-feature construction, soft-token generation and export, and evidence retrieval
and packing. These measurements include I/O within each stage, while LLM inference is evaluated separately.

\begin{figure}[t]
  \centering
  \includegraphics[width=0.7\linewidth]{\detokenize{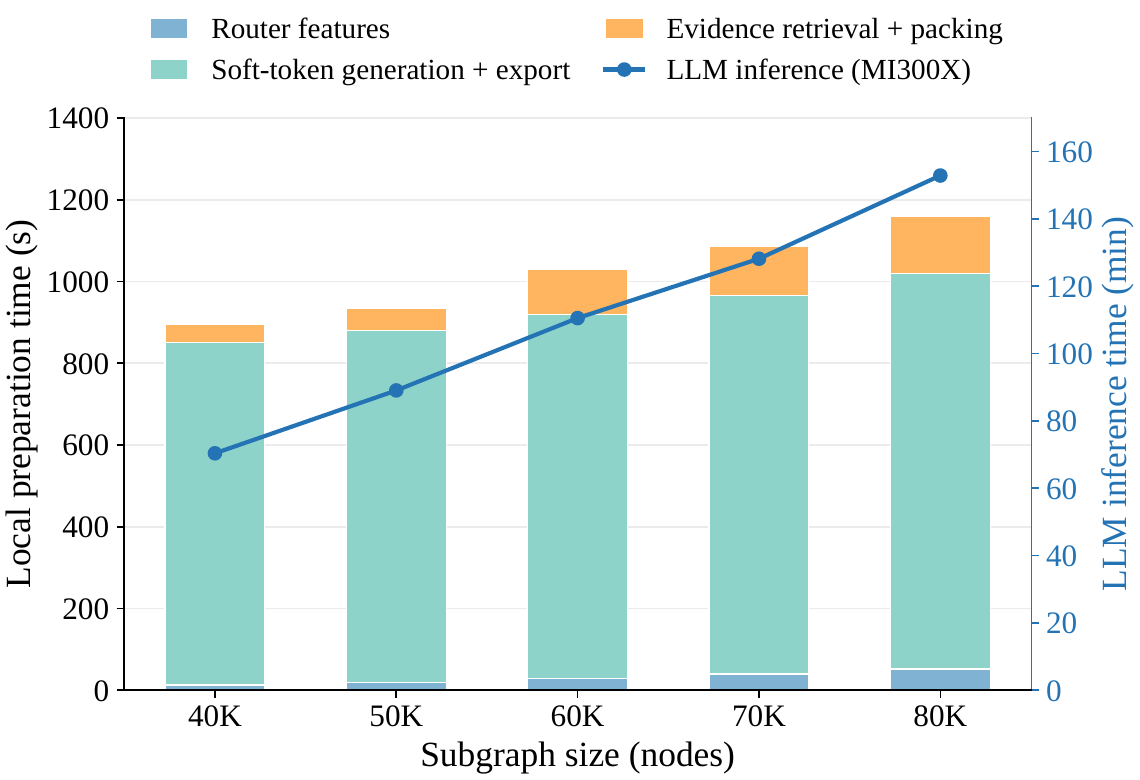}}
  \caption{Local preparation time and inference time on Ele-fashion.
  }
  \label{fig:ele-local-preparation}
\vspace{-0.1cm}
\end{figure}

As shown in Fig.~\ref{fig:ele-local-preparation}, preparation time
increases gradually with graph size, with soft-token generation and export accounting for the largest share.
LLM inference time slightly nearly doubles as the node count increases from 40K to 80K.
These results indicate manageable runtime growth for both preparation and inference over the evaluated range.

\subsection{Meta-path Routing Analysis (RQ5)}
\label{sec:routing_analysis}
\begin{figure*}[t]
    \centering
    \includegraphics[width=\textwidth]{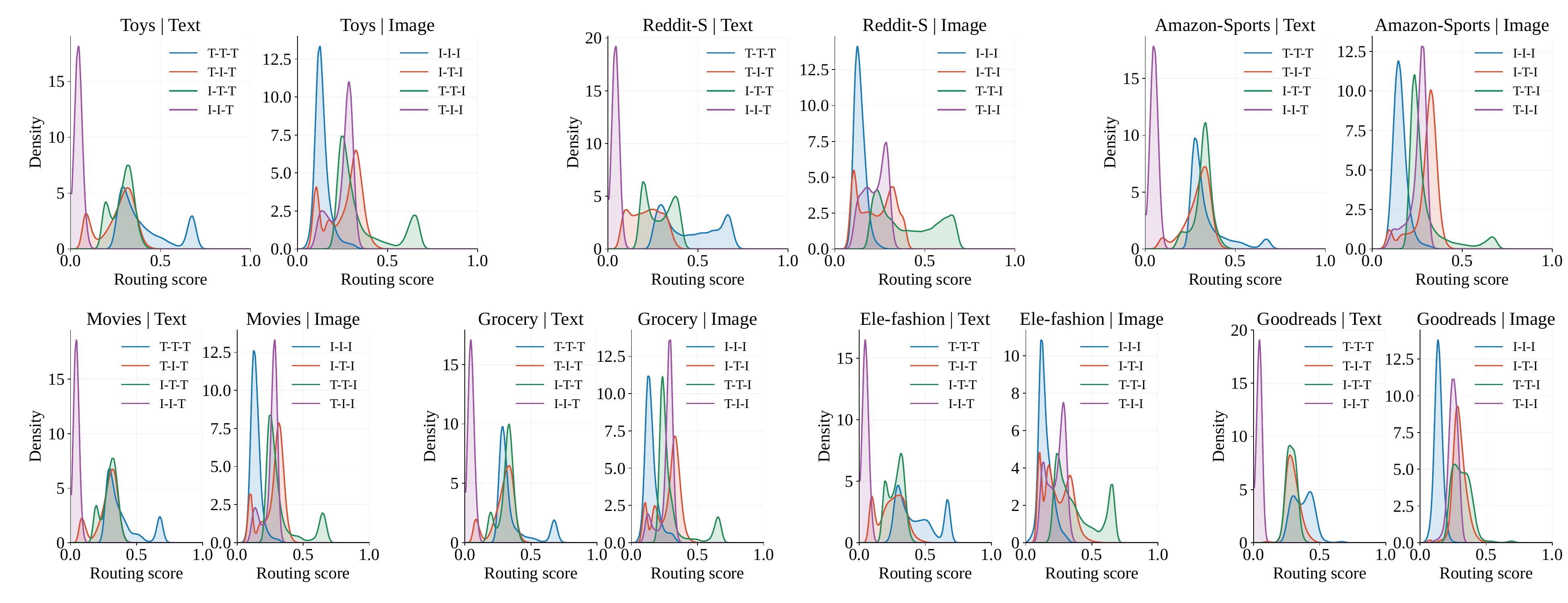}
    \caption{Meta-path routing-score distributions from the final
    HMP block across seven datasets.
    The top and bottom rows show source and target datasets,
    respectively. Each dataset contains paired text-branch
    (left) and image-branch (right) panels, with four meta-paths
    per branch. 
    }
    \label{fig:routing_distributions}
\vspace{-0.0cm}
\end{figure*}

\noindent\textbf{Dataset-level routing score distribution.}
Figure~\ref{fig:routing_distributions} visualizes the node-wise
routing-score distributions of the four text-ending and four
image-ending meta-paths. These distributions reveal both shared
routing patterns across related datasets and dataset-dependent
variations in path weighting.

\noindent\textit{Differences among source datasets.}
The three source datasets exhibit distinct routing profiles.
Despite their common Amazon origin, Toys (NC) and Amazon-Sports
(LP) differ in the image-ending path $T\rightarrow T\rightarrow I$
(green curve): its scores are more concentrated around $0.2$--$0.3$
on Amazon-Sports, whereas Toys exhibits a more noticeable secondary
mode around $0.6$--$0.7$. Reddit-S shows a broader distribution
for $T\rightarrow T\rightarrow I$ (green curve, image branch)
and a more multimodal distribution for $I\rightarrow T\rightarrow T$
(green curve, text branch).
These observations indicate that a shared platform does not imply
identical routing behavior.

\noindent\textit{Similarities across related datasets.}
Toys, Movies, and Grocery, all Amazon-based NC datasets, display
qualitatively similar routing patterns. In the text branch,
$I\rightarrow I\rightarrow T$ (purple curve) receives consistently
low weights, while $T\rightarrow T\rightarrow T$ (blue curve)
retains a high-score secondary mode around $0.6$--$0.7$.
In the image branch, $I\rightarrow I\rightarrow I$ (blue curve)
is concentrated at relatively low scores, whereas
$T\rightarrow I\rightarrow I$ (purple curve) and
$I\rightarrow T\rightarrow I$ (orange-red curve) exhibit their
main modes around $0.25$--$0.35$.
These similarities suggest common propagation preferences across
related graph contexts.

\noindent\textit{Variations across target datasets.}
Ele-fashion and Goodreads exhibit different routing profiles.
In the text branch, Ele-fashion retains a noticeable high-score
mode near $0.7$ for $T\rightarrow T\rightarrow T$ (blue curve),
which is absent from Goodreads.
In the image branch, $T\rightarrow T\rightarrow I$ (green curve)
has a broader distribution on Ele-fashion, with a secondary mode
near $0.65$, whereas its scores on Goodreads concentrate mainly
around $0.25$--$0.4$.
Overall, these patterns illustrate that MSPR does not impose a
single fixed weighting configuration across datasets, but produces
different distributions of meta-path preferences in different
graph contexts.

\begin{figure}[t]
  \centering
  \begin{tabular}{@{}c@{}c@{}}
    \multicolumn{2}{@{}c@{}}{%
      \includegraphics[width=\linewidth]
      {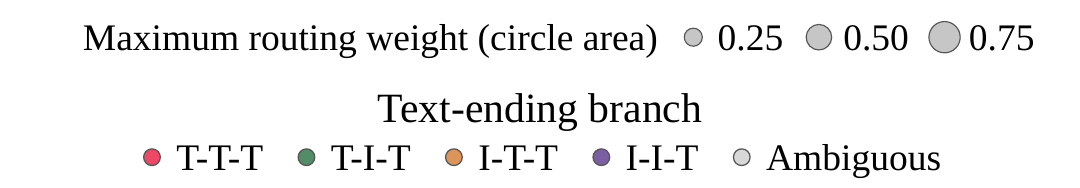}%
    }\\[-2pt]

    \includegraphics[width=0.49\linewidth]
      {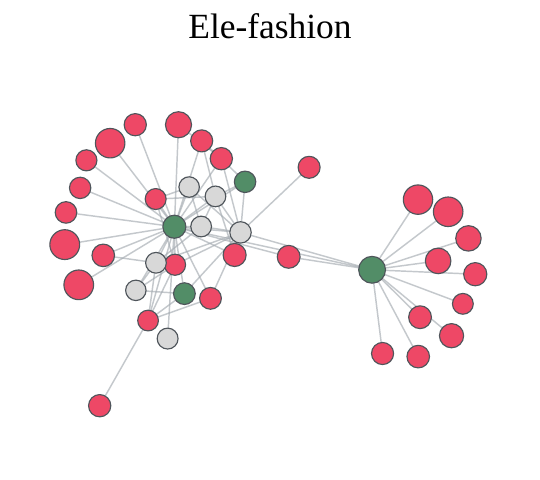} &
    \includegraphics[width=0.49\linewidth]
      {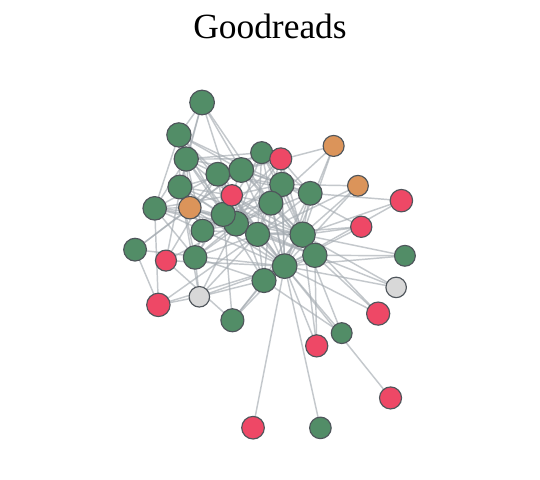}
    \\[-2pt]

    \multicolumn{2}{@{}c@{}}{%
      \includegraphics[width=\linewidth]
      {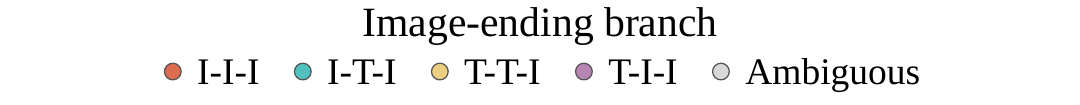}%
    }\\[-2pt]

    \includegraphics[width=0.49\linewidth]
      {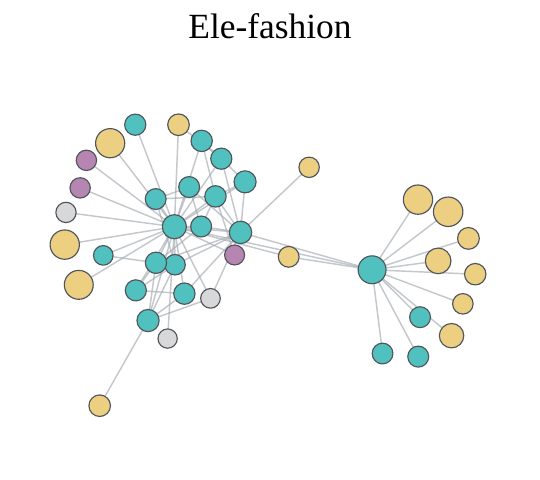} &
    \includegraphics[width=0.49\linewidth]
      {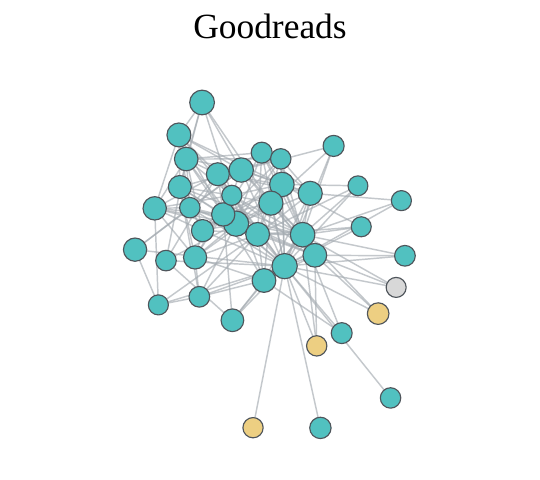}
  \end{tabular}
  \caption{Node-level meta-path routing preferences on
  Ele-fashion (left) and Goodreads (right).
  Gray denotes a gap of at most 0.01 between the two largest
  weights.
  }
  \label{fig:node-routing-preferences}
\vspace{-0.0cm}
\end{figure}


\noindent\textbf{Node-level Meta-path Preferences.}
Fig.~\ref{fig:node-routing-preferences} visualizes the highest-weight meta-paths in sampled local subgraphs from four datasets.
The text-terminal branch exhibits distinct preferences across these subgraphs. 
Ele-fashion is dominated by T-T-T (rose red), while Goodreads
mainly favors T-I-T (green).

The image-terminal branch exhibits a more consistent dominant pattern:
I-T-I (turquoise) is the most frequent preference in these twpo subgraphs and is preferred by every displayed Grocery node.
Nevertheless, node-level variation remains visible.
Ele-fashion also contain nodes favoring T-T-I (yellow)
or T-I-I (mauve), whereas only a few displayed Goodreads nodes
favor T-T-I (yellow).
Thus, a shared dominant meta-path can coexist with heterogeneous
preferences among individual nodes.

Importantly, the dominant paths in the two terminal-modality branches
do not follow a fixed one-to-one correspondence.
For example, the nodes on the right side of the Ele-fashion subgraph
consistently favor T-T-T (rose red) in the text-terminal branch,
but their image-terminal preferences differ between T-T-I (yellow)
and I-T-I (turquoise).
Similarly, many Goodreads nodes share an I-T-I (turquoise)
preference in the image-terminal branch while exhibiting different
text-terminal preferences, including T-I-T (green) and
T-T-T (rose red).
These examples illustrate that routing preferences vary not only
across nodes and datasets, but also between terminal modalities
for the same nodes.

\section{Conclusion}
In this paper, we proposed MAGE, a framework for transferable multimodal graph learning. MAGE explicitly models cross-node and cross-modal interactions through heterogeneous meta-paths
and adaptively aggregates textual and visual neighborhood information using node-specific routing. Evidence-guided instruction tuning then incorporates the resulting graph-aware
representations of targets and selected evidence into a frozen LLM. Through joint training on source-domain node classification and link prediction tasks, MAGE supports prediction on unseen
target graphs without target-domain parameter updates.
Experiments demonstrate consistent improvements in node classification across four target datasets. Routing analyses further reveal distinct meta-path preferences across datasets,
nodes, and terminal modalities. These findings support adaptive multimodal neighborhood aggregation and graph-aware evidence as an effective approach to cross-domain graph prediction.



\bibliographystyle{IEEEtran}
\bibliography{sample-base}

@String{Computing = "Computing" }

@String{Computer = "{IEEE} Computer" }

@String{Springer = "Springer-Verlag" }

@ArtifactSoftware{R,
    title = {R: A Language and Environment for Statistical Computing},
    author = {{R Core Team}},
    organization = {R Foundation for Statistical Computing},
    address = {Vienna, Austria},
    year = {2019},
    url = {https://www.R-project.org/},
}

@inproceedings{he2025unigraph2,
  title={Unigraph2: Learning a unified embedding space to bind multimodal graphs},
  author={He, Yufei and Sui, Yuan and He, Xiaoxin and Liu, Yue and Sun, Yifei and Hooi, Bryan},
  booktitle={Proceedings of the ACM on Web Conference 2025},
  pages={1759--1770},
  year={2025}
}

@inproceedings{hong2026multimodal,
  title={Multimodal graph representation learning with dynamic information pathways},
  author={Hong, Xiaobin and Lin, Mingkai and Wang, Xiaoli and Wang, Chaoqun and Li, Wenzhong},
  booktitle={Proceedings of the AAAI Conference on Artificial Intelligence},
  volume={40},
  number={17},
  pages={14829--14837},
  year={2026}
}

@inproceedings{zhang2025multiple,
  title={Multiple heads are better than one: Mixture of modality knowledge experts for entity representation learning},
  author={Zhang, Yichi and Chen, Zhuo and Guo, Lingbing and Hu, Binbin and Liu, Ziqi and Zhang, Wen and Chen, Huajun and others},
  booktitle={International Conference on Learning Representations},
  volume={2025},
  pages={54811--54828},
  year={2025}
}

@inproceedings{peng2025modalities,
  title={Modalities Contribute Unequally: Enhancing Medical Multi-modal Learning through Adaptive Modality Token Re-balancing.},
  author={Peng, Jie and Ballard, Jenna L and Zhang, Mohan and Yun, Sukwon and Xin, Jiayi and Long, Qi and Zhang, Yanyong and Chen, Tianlong},
  booktitle={ICML},
  year={2025}
}

@article{sun2026mario,
  title={Mario: Multimodal graph reasoning with large language models},
  author={Sun, Yuanfu and Li, Kang and Guo, Pengkang and Liu, Jiajin and Tan, Qiaoyu},
  journal={arXiv preprint arXiv:2603.05181},
  year={2026}
}

@inproceedings{fan2026mlaga,
  title = {{MLaGA}: Multimodal Large Language and Graph Assistant},
  author = {Fan, Dongzhe and Liu, Jiajin and Fang, Yi and Difallah, Djellel and Tan, Qiaoyu},
  booktitle = {Proceedings of the 32nd ACM SIGKDD Conference on Knowledge Discovery and Data Mining V. 2},
  pages = {1027--1038},
  year = {2026},
  publisher = {Association for Computing Machinery},
  address = {New York, NY, USA}
}

@inproceedings{tang2024graphgpt,
  title={Graphgpt: Graph instruction tuning for large language models},
  author={Tang, Jiabin and Yang, Yuhao and Wei, Wei and Shi, Lei and Su, Lixin and Cheng, Suqi and Yin, Dawei and Huang, Chao},
  booktitle={Proceedings of the 47th international ACM SIGIR conference on research and development in information retrieval},
  pages={491--500},
  year={2024}
}

@inproceedings{liu2024can,
  title={Can we soft prompt llms for graph learning tasks?},
  author={Liu, Zheyuan and He, Xiaoxin and Tian, Yijun and Chawla, Nitesh V},
  booktitle={Companion proceedings of the ACM web conference 2024},
  pages={481--484},
  year={2024}
}

@article{chen2024llaga,
  title={Llaga: Large language and graph assistant},
  author={Chen, Runjin and Zhao, Tong and Jaiswal, Ajay and Shah, Neil and Wang, Zhangyang},
  journal={arXiv preprint arXiv:2402.08170},
  year={2024}
}

@inproceedings{zhang2024graphtranslator,
  title={Graphtranslator: Aligning graph model to large language model for open-ended tasks},
  author={Zhang, Mengmei and Sun, Mingwei and Wang, Peng and Fan, Shen and Mo, Yanhu and Xu, Xiaoxiao and Liu, Hong and Yang, Cheng and Shi, Chuan},
  booktitle={Proceedings of the ACM Web Conference 2024},
  pages={1003--1014},
  year={2024}
}

@article{wang2024llms,
  title={Llms as zero-shot graph learners: Alignment of gnn representations with llm token embeddings},
  author={Wang, Duo and Zuo, Yuan and Li, Fengzhi and Wu, Junjie},
  journal={Advances in neural information processing systems},
  volume={37},
  pages={5950--5973},
  year={2024}
}

@inproceedings{kong2025gofa,
  title={Gofa: A generative one-for-all model for joint graph language modeling},
  author={Kong, Lecheng and Feng, Jiarui and Liu, Hao and Huang, Chengsong and Huang, Jiaxin and Chen, Yixin and Zhang, Muhan},
  booktitle={International Conference on Learning Representations},
  volume={2025},
  pages={40792--40823},
  year={2025}
}

@inproceedings{li2023blip,
  title={Blip-2: Bootstrapping language-image pre-training with frozen image encoders and large language models},
  author={Li, Junnan and Li, Dongxu and Savarese, Silvio and Hoi, Steven},
  booktitle={International conference on machine learning},
  pages={19730--19742},
  year={2023},
  organization={PmLR}
}

@inproceedings{radford2021learning,
  title={Learning transferable visual models from natural language supervision},
  author={Radford, Alec and Kim, Jong Wook and Hallacy, Chris and Ramesh, Aditya and Goh, Gabriel and Agarwal, Sandhini and Sastry, Girish and Askell, Amanda and Mishkin, Pamela and Clark, Jack and others},
  booktitle={International conference on machine learning},
  pages={8748--8763},
  year={2021},
  organization={PmLR}
}

@article{liu2023visual,
  title={Visual instruction tuning},
  author={Liu, Haotian and Li, Chunyuan and Wu, Qingyang and Lee, Yong Jae},
  journal={Advances in neural information processing systems},
  volume={36},
  pages={34892--34916},
  year={2023}
}

@article{zheng2023judging,
  title={Judging llm-as-a-judge with mt-bench and chatbot arena},
  author={Zheng, Lianmin and Chiang, Wei-Lin and Sheng, Ying and Zhuang, Siyuan and Wu, Zhanghao and Zhuang, Yonghao and Lin, Zi and Li, Zhuohan and Li, Dacheng and Xing, Eric and others},
  journal={Advances in neural information processing systems},
  volume={36},
  pages={46595--46623},
  year={2023}
}

@inproceedings{liu2024improved,
  title={Improved baselines with visual instruction tuning},
  author={Liu, Haotian and Li, Chunyuan and Li, Yuheng and Lee, Yong Jae},
  booktitle={2024 IEEE/CVF Conference on Computer Vision and Pattern Recognition (CVPR)},
  pages={26286--26296},
  year={2024},
  organization={IEEE}
}

@article{bai2025qwen25vl,
  title={{Qwen2.5-VL} Technical Report},
  author={Bai, Shuai and Chen, Keqin and Liu, Xuejing and others},
  journal={arXiv preprint arXiv:2502.13923},
  year={2025},
  url={https://arxiv.org/abs/2502.13923}
}

@inproceedings{ni2019justifying,
  title={Justifying recommendations using distantly-labeled reviews and fine-grained aspects},
  author={Ni, Jianmo and Li, Jiacheng and McAuley, Julian},
  booktitle={Proceedings of the 2019 conference on empirical methods in natural language processing and the 9th international joint conference on natural language processing (EMNLP-IJCNLP)},
  pages={188--197},
  year={2019}
}

@article{yan2024graph,
  title={When graph meets multimodal: Benchmarking and meditating on multimodal attributed graphs learning},
  author={Yan, Hao and Li, Chaozhuo and Yin, Jun and Yu, Zhigang and Han, Weihao and Li, Mingzheng and Zeng, Zhengxin and Sun, Hao and Wang, Senzhang},
  journal={arXiv preprint arXiv:2410.09132},
  year={2024}
}

@article{desai2021redcaps,
  title={Redcaps: Web-curated image-text data created by the people, for the people},
  author={Desai, Karan and Kaul, Gaurav and Aysola, Zubin and Johnson, Justin},
  journal={arXiv preprint arXiv:2111.11431},
  year={2021}
}

@inproceedings{wan2018item,
  title={Item recommendation on monotonic behavior chains},
  author={Wan, Mengting and McAuley, Julian},
  booktitle={Proceedings of the 12th ACM conference on recommender systems},
  pages={86--94},
  year={2018}
}

@inproceedings{zhu2025mosaic,
  title={Mosaic of modalities: A comprehensive benchmark for multimodal graph learning},
  author={Zhu, Jing and Zhou, Yuhang and Qian, Shengyi and He, Zhongmou and Zhao, Tong and Shah, Neil and Koutra, Danai},
  booktitle={2025 IEEE/CVF Conference on Computer Vision and Pattern Recognition (CVPR)},
  pages={14215--14224},
  year={2025},
  organization={IEEE}
}

@article{sun2011pathsim,
  title={Pathsim: Meta path-based top-k similarity search in heterogeneous information networks},
  author={Sun, Yizhou and Han, Jiawei and Yan, Xifeng and Yu, Philip S and Wu, Tianyi},
  journal={Proceedings of the VLDB Endowment},
  volume={4},
  number={11},
  pages={992--1003},
  year={2011},
  publisher={VLDB Endowment}
}

@inproceedings{li2021leveraging,
  title={Leveraging meta-path contexts for classification in heterogeneous information networks},
  author={Li, Xiang and Ding, Danhao and Kao, Ben and Sun, Yizhou and Mamoulis, Nikos},
  booktitle={2021 IEEE 37th International Conference on Data Engineering (ICDE)},
  pages={912--923},
  year={2021},
  organization={IEEE}
}

@inproceedings{liang2024hgamlp,
  title={Hgamlp: Heterogeneous graph attention mlp with de-redundancy mechanism},
  author={Liang, Yuxuan and Zhang, Wentao and Sheng, Zeang and Yang, Ling and Jiang, Jiawei and Tong, Yunhai and Cui, Bin},
  booktitle={2024 IEEE 40th International Conference on Data Engineering (ICDE)},
  pages={2779--2791},
  year={2024},
  organization={IEEE}
}

@inproceedings{zhang2025effective,
  title={Effective and Scalable Heterogeneous Graph Neural Network Framework with Convolution-oriented Attention},
  author={Zhang, Ziqian and Wang, Chaokun and Zheng, Shuwen and Wu, Cheng and Liu, Ziyang and Feng, Hao},
  booktitle={2025 IEEE 41st International Conference on Data Engineering (ICDE)},
  pages={4011--4024},
  year={2025},
  organization={IEEE}
}

@article{liu2024msgnn,
  title={MSGNN: Masked Schema based Graph Neural Networks},
  author={Liu, Hao and Yang, Qianwen and Cui, Taoyong and Wang, Wei},
  journal={Proceedings of the VLDB Endowment},
  volume={18},
  number={3},
  pages={571--584},
  year={2024},
  publisher={VLDB Endowment}
}

@inproceedings{shan2026mug,
  title={MUG: Meta-path-aware Universal Heterogeneous Graph Pre-Training},
  author={Shan, Lianze and Zhao, Jitao and He, Dongxiao and Huang, Yongqi and Feng, Zhiyong and Zhang, Weixiong},
  booktitle={Proceedings of the AAAI Conference on Artificial Intelligence},
  volume={40},
  number={30},
  pages={25260--25268},
  year={2026}
}

@inproceedings{deng2023construction,
  title={Construction and applications of billion-scale pre-trained multimodal business knowledge graph},
  author={Deng, Shumin and Wang, Chengming and Li, Zhoubo and Zhang, Ningyu and Dai, Zelin and Chen, Hehong and Xiong, Feiyu and Yan, Ming and Chen, Qiang and Chen, Mosha and others},
  booktitle={2023 IEEE 39th International Conference on Data Engineering (ICDE)},
  pages={2988--3002},
  year={2023},
  organization={IEEE}
}

@inproceedings{he2024firzen,
  title={Firzen: Firing strict cold-start items with frozen heterogeneous and homogeneous graphs for recommendation},
  author={He, Hulingxiao and He, Xiangteng and Peng, Yuxin and Shan, Zifei and Su, Xin},
  booktitle={2024 IEEE 40th international conference on data engineering (ICDE)},
  pages={4657--4670},
  year={2024},
  organization={IEEE}
}

@inproceedings{chen2025birds,
  title={Birds of a Feather: Enhancing Multimodal Fake News Detection via Multi-Element Retrieval},
  author={Chen, Xueqin and Huang, Xiaoyu and Gao, Qiang and Huang, Li and Yu, Jiajing and Liu, Guisong},
  booktitle={2025 IEEE 41st International Conference on Data Engineering (ICDE)},
  pages={1--14},
  year={2025},
  organization={IEEE}
}

@inproceedings{cheng2024retrieval,
  title={Retrieval-augmented hypergraph for multimodal social media popularity prediction},
  author={Cheng, Zhangtao and Zhang, Jienan and Xu, Xovee and Trajcevski, Goce and Zhong, Ting and Zhou, Fan},
  booktitle={Proceedings of the 30th ACM SIGKDD conference on knowledge discovery and data mining},
  pages={445--455},
  year={2024}
}

@inproceedings{zhou2025multimodal,
  title={Multimodal graph-based variational mixture of experts network for zero-shot multimodal information extraction},
  author={Zhou, Baohang and Zhang, Ying and Zhao, Yu and Sui, Xuhui and Yuan, Xiaojie},
  booktitle={Proceedings of the ACM on Web Conference 2025},
  pages={4823--4831},
  year={2025}
}

@article{li2026modalities,
  title={When modalities remember: Continual learning for multimodal knowledge graphs},
  author={Li, Linyu and Jin, Zhi and Zhang, Yichi and Jin, Dongming and He, Yuanpeng and Duan, Haoran and Luosang, Gadeng and Tashi, Nyima},
  journal={arXiv preprint arXiv:2604.02778},
  year={2026}
}

@article{raffel2020exploring,
  title={Exploring the limits of transfer learning with a unified text-to-text transformer},
  author={Raffel, Colin and Shazeer, Noam and Roberts, Adam and Lee, Katherine and Narang, Sharan and Matena, Michael and Zhou, Yanqi and Li, Wei and Liu, Peter J},
  journal={Journal of machine learning research},
  volume={21},
  number={140},
  pages={1--67},
  year={2020}
}

@article{dosovitskiy2020image,
  title={An image is worth 16x16 words: Transformers for image recognition at scale},
  author={Dosovitskiy, Alexey},
  journal={arXiv preprint arXiv:2010.11929},
  year={2020}
}

@article{kipf2016semi,
  title={Semi-supervised classification with graph convolutional networks},
  author={Kipf, TN},
  journal={arXiv preprint arXiv:1609.02907},
  year={2016}
}

@article{kang2026ss,
  title={Ss-adamoe: Spatio-spectral adaptive mixture of experts with global structural priors for graph node classification},
  author={Kang, Xilin and Yu, Tianyue and Wang, Letao and Guo, Yutong and Zhang, Fengjun},
  journal={Entropy},
  volume={28},
  number={3},
  pages={355},
  year={2026},
  publisher={MDPI}
}

@article{ma2024mixture,
  title={Mixture of link predictors on graphs},
  author={Ma, Li and Han, Haoyu and Li, Juanhui and Shomer, Harry and Liu, Hui and Gao, Xiaofeng and Tang, Jiliang},
  journal={Advances in Neural Information Processing Systems},
  volume={37},
  pages={16043--16070},
  year={2024}
}

@inproceedings{wang2019heterogeneous,
  title={Heterogeneous graph attention network},
  author={Wang, Xiao and Ji, Houye and Shi, Chuan and Wang, Bai and Ye, Yanfang and Cui, Peng and Yu, Philip S},
  booktitle={The world wide web conference},
  pages={2022--2032},
  year={2019}
}

@inproceedings{fu2020magnn,
  title={Magnn: Metapath aggregated graph neural network for heterogeneous graph embedding},
  author={Fu, Xinyu and Zhang, Jiani and Meng, Ziqiao and King, Irwin},
  booktitle={Proceedings of the web conference 2020},
  pages={2331--2341},
  year={2020}
}

@inproceedings{hu2020heterogeneous,
  title={Heterogeneous graph transformer},
  author={Hu, Ziniu and Dong, Yuxiao and Wang, Kuansan and Sun, Yizhou},
  booktitle={Proceedings of the web conference 2020},
  pages={2704--2710},
  year={2020}
}

@inproceedings{wang2021self,
  title={Self-supervised heterogeneous graph neural network with co-contrastive learning},
  author={Wang, Xiao and Liu, Nian and Han, Hui and Shi, Chuan},
  booktitle={Proceedings of the 27th ACM SIGKDD conference on knowledge discovery \& data mining},
  pages={1726--1736},
  year={2021}
}

@inproceedings{qiao2025login,
  title={Login: A large language model consulted graph neural network training framework},
  author={Qiao, Yiran and Ao, Xiang and Liu, Yang and Xu, Jiarong and Sun, Xiaoqian and He, Qing},
  booktitle={Proceedings of the Eighteenth ACM International Conference on Web Search and Data Mining},
  pages={232--241},
  year={2025}
}

@inproceedings{luo2025graph,
  title={Graph-based cross-domain knowledge distillation for cross-dataset text-to-image person retrieval},
  author={Luo, Bingjun and Wang, Jinpeng and Wang, Zewen and Zhu, Junjie and Zhao, Xibin},
  booktitle={Proceedings of the AAAI Conference on Artificial Intelligence},
  volume={39},
  number={1},
  pages={568--576},
  year={2025}
}

@inproceedings{wang2024instructgraph,
  title={Instructgraph: Boosting large language models via graph-centric instruction tuning and preference alignment},
  author={Wang, Jianing and Wu, Junda and Hou, Yupeng and Liu, Yao and Gao, Ming and McAuley, Julian},
  booktitle={Findings of the Association for Computational Linguistics: ACL 2024},
  pages={13492--13510},
  year={2024}
}

@inproceedings{he2024harnessing,
  title={Harnessing explanations: Llm-to-lm interpreter for enhanced text-attributed graph representation learning},
  author={He, Xiaoxin and Bresson, Xavier and Laurent, Thomas and Perold, Adam and LeCun, Yann and Hooi, Bryan},
  booktitle={International conference on learning representations},
  volume={2024},
  pages={5711--5732},
  year={2024}
}

@article{li2024similarity,
  title={Similarity-based neighbor selection for graph llms},
  author={Li, Rui and Li, Jiwei and Han, Jiawei and Wang, Guoyin},
  journal={arXiv preprint arXiv:2402.03720},
  year={2024}
}

@article{he2024g,
  title={G-retriever: Retrieval-augmented generation for textual graph understanding and question answering},
  author={He, Xiaoxin and Tian, Yijun and Sun, Yifei and Chawla, Nitesh V and Laurent, Thomas and LeCun, Yann and Bresson, Xavier and Hooi, Bryan},
  journal={Advances in Neural Information Processing Systems},
  volume={37},
  pages={132876--132907},
  year={2024}
}

@article{mortaheb2025re,
  title={Re-ranking the context for multimodal retrieval augmented generation},
  author={Mortaheb, Matin and Khojastepour, Mohammad A Amir and Chakradhar, Srimat T and Ulukus, Sennur},
  journal={arXiv preprint arXiv:2501.04695},
  year={2025}
}

@article{shu2020fakenewsnet,
  title={Fakenewsnet: A data repository with news content, social context, and spatiotemporal information for studying fake news on social media},
  author={Shu, Kai and Mahudeswaran, Deepak and Wang, Suhang and Lee, Dongwon and Liu, Huan},
  journal={Big data},
  volume={8},
  number={3},
  pages={171--188},
  year={2020},
  publisher={SAGE Publications, Sage CA: Los Angeles, CA}
}

@inproceedings{qi2012community,
  title={Community detection with edge content in social media networks},
  author={Qi, Guo-Jun and Aggarwal, Charu C and Huang, Thomas},
  booktitle={2012 IEEE 28th International conference on data engineering},
  pages={534--545},
  year={2012},
  organization={IEEE}
}

@inproceedings{mcauley2015image,
  title={Image-based recommendations on styles and substitutes},
  author={McAuley, Julian and Targett, Christopher and Shi, Qinfeng and Van Den Hengel, Anton},
  booktitle={Proceedings of the 38th international ACM SIGIR conference on research and development in information retrieval},
  pages={43--52},
  year={2015}
}

@inproceedings{zhang2021mining,
  title={Mining latent structures for multimedia recommendation},
  author={Zhang, Jinghao and Zhu, Yanqiao and Liu, Qiang and Wu, Shu and Wang, Shuhui and Wang, Liang},
  booktitle={Proceedings of the 29th ACM international conference on multimedia},
  pages={3872--3880},
  year={2021}
}

@article{li2026choe,
  title={CHoE: Cross-Domain Heterogeneous Graph Prompt Learning via Structure-Conditioned Experts},
  author={Li, Peiyuan and Huang, Yongqi and Zhao, Jitao and He, Dongxiao and Jin, Di and Zhang, Weixiong},
  journal={arXiv preprint arXiv:2605.15888},
  year={2026}
}

@article{zhu2021transfer,
  title={Transfer learning of graph neural networks with ego-graph information maximization},
  author={Zhu, Qi and Yang, Carl and Xu, Yidan and Wang, Haonan and Zhang, Chao and Han, Jiawei},
  journal={Advances in Neural Information Processing Systems},
  volume={34},
  pages={1766--1779},
  year={2021}
}

@article{yarici2026mer,
  title={MER-DG: Modality-Entropy Regularization for Multimodal Domain Generalization},
  author={Yarici, Yavuz and AlRegib, Ghassan},
  journal={arXiv preprint arXiv:2605.01967},
  year={2026}
}

@article{planamente2024relative,
  title={Relative norm alignment for tackling domain shift in deep multi-modal classification},
  author={Planamente, Mirco and Plizzari, Chiara and Peirone, Simone Alberto and Caputo, Barbara and Bottino, Andrea},
  journal={International Journal of Computer Vision},
  volume={132},
  number={7},
  pages={2618--2638},
  year={2024},
  publisher={Springer}
}

@article{fan2024cross,
  title={Cross-modal representation flattening for multi-modal domain generalization},
  author={Fan, Yunfeng and Xu, Wenchao and Wang, Haohao and Guo, Song},
  journal={Advances in Neural Information Processing Systems},
  volume={37},
  pages={66773--66795},
  year={2024}
}

@article{dai2023instructblip,
  title={Instructblip: Towards general-purpose vision-language models with instruction tuning},
  author={Dai, Wenliang and Li, Junnan and Li, Dongxu and Tiong, Anthony and Zhao, Junqi and Wang, Weisheng and Li, Boyang and Fung, Pascale N and Hoi, Steven},
  journal={Advances in neural information processing systems},
  volume={36},
  pages={49250--49267},
  year={2023}
}

@article{alayrac2022flamingo,
  title={Flamingo: a visual language model for few-shot learning},
  author={Alayrac, Jean-Baptiste and Donahue, Jeff and Luc, Pauline and Miech, Antoine and Barr, Iain and Hasson, Yana and Lenc, Karel and Mensch, Arthur and Millican, Katherine and Reynolds, Malcolm and others},
  journal={Advances in neural information processing systems},
  volume={35},
  pages={23716--23736},
  year={2022}
}

@inproceedings{zhai2023sigmoid,
  title={Sigmoid loss for language image pre-training},
  author={Zhai, Xiaohua and Mustafa, Basil and Kolesnikov, Alexander and Beyer, Lucas},
  booktitle={2023 IEEE/CVF International Conference on Computer Vision (ICCV)},
  pages={11941--11952},
  year={2023},
  organization={IEEE}
}

\end{document}